\documentclass[a4paper,fleqn]{cas-dc}
\usepackage{cite}
\usepackage{amsmath,amssymb,amsfonts}
\usepackage{algorithmic}
\usepackage{graphicx}
\usepackage{textcomp}
\usepackage{xcolor}
\usepackage{subfigure}

\usepackage[numbers]{natbib}

\usepackage[switch]{lineno} 

\def\tsc#1{\csdef{#1}{\textsc{\lowercase{#1}}\xspace}}
\tsc{WGM}
\tsc{QE}

\begin{document}
\let\WriteBookmarks\relax
\def\floatpagepagefraction{1}
\def\textpagefraction{.001}

\shorttitle{Comparing Global Tourism Flows Measured by Official Census and Social Sensing}    

\shortauthors{Skora, Silva, Senefonte, Delgado and Lüders}  

\title [mode = title]{Comparing Global Tourism Flows Measured by Official Census and Social Sensing}  

\author[1]{Lucas E. B. Skora}[
    orcid=0000-0002-9585-496X,
]
\ead{lucasskora@alunos.utfpr.edu.br}

\author[1, 2]{Helen C. M. Senefonte}[
    orcid=0000-0001-8990-9162,
]
\ead{helen@uel.br}

\author[1]{Myriam Regattieri Delgado}[
    orcid=0000-0002-2791-174X,
]
\ead{myriamdelg@utfpr.edu.br}

\author[1]{Ricardo L\"uders}[
    orcid=0000-0001-6483-4694 ,
]
\ead{luders@utfpr.edu.br}

\author[1, 3]{Thiago H. Silva}[
    orcid=0000-0001-6994-8076,
]
\ead{thiagoh@utfpr.edu.br}


\affiliation[1]{
    organization={Universidade Tecnológica Federal do Paraná (UTFPR)},
    addressline={Av. Sete de Setembro, 3165}, 
    city={Curitiba},
    postcode={80230-901}, 
    country={Brazil}
}

\affiliation[2]{
    organization={Universidade Estadual de Londrina (UEL)},
    addressline={Rodovia Celso Garcia Cid, PR-445, Km 380}, 
    city={Londrina},
    postcode={86057-970}, 
    country={Brazil}
}

\affiliation[3]{
    organization={University of Toronto},
    addressline={55 St. George Street}, 
    city={Toronto},
    postcode={M5S 0C9}, 
    country={Canada}}

\begin{abstract}
A better understanding of the behavior of tourists is strategic for improving services in the competitive and important economic segment of global tourism. Critical studies in the literature often explore the issue using traditional data, such as questionnaires or interviews. Traditional approaches provide precious information; however, they impose challenges to obtaining large-scale data, making it hard to study worldwide patterns. Location-based social networks (LBSNs) can potentially mitigate such issues due to the relatively low cost of acquiring large amounts of behavioral data. Nevertheless, before using such data for studying tourists' behavior, it is necessary to verify whether the information adequately reveals the behavior measured with traditional data -- considered the ground truth. Thus, the present work investigates in which countries the global tourism network measured with an LBSN agreeably reflects the behavior estimated by the World Tourism Organization using traditional methods. Although we could find exceptions, the results suggest that, for most countries, LBSN data can satisfactorily represent the behavior studied. We have an indication that, in countries with high correlations between results obtained from both datasets, LBSN data can be used in research regarding the mobility of the tourists in the studied context.
\end{abstract}


\maketitle

\section{Introduction}

Several studies available in the literature explore tourist behavior in different types of data sources. This type of study is strategic from an economic and social standpoint \cite{unwto20}. Many of these works use traditional survey data \cite{Zieba17, Scuderi18}, which are normally the ground truths for the various questions they address. For example, Lozano and Guti{\' e}rrez \cite{lozano2018complex} study the global tourism network using data from the World Tourism Organization (WTO) obtained in a traditional way, leading to important insights for the development of new strategies concerning international tourism.

However, official data such as those from  WTO are typically difficult to obtain because traditional approaches to data collection do not scale easily. Trying to solve this problem, recent studies are using data from location-based social networks (LBSNs) \cite{zheng2014, Silva:2019:UCL:3309872.3301284}. LBSN data scale more easily and allow analysis at different granularities, going down to group or even individual level in specific areas. Traditional sources generally provide aggregated results for some particular locations, such as countries or specific cities \cite{Silva:2019:UCL:3309872.3301284}, but usually do not reach the specificity of social networks. As such, data from these new sources have the potential to complement traditional ones in different ways.

Distinct results are observed when using similar methods in different datasets, especially when datasets have peculiar characteristics and origins, such as social media data. Research exploring the link between traditional survey and social media data can improve studies of social phenomena \cite{murphy2014} and alleviate some methodological problems of research based on only one of the types of data sources \cite{baghal2021}.

Validating LSBN data becomes essential to explore the potential for their use in the study of tourist mobility. Therefore, the main research question that guides this work is:  Where does LBSN data accurately represent the global tourism network compared with traditional data provided by WTO? 

To answer this question, we analyze differences and similarities between information obtained from the official WTO data explored by Lozano and Guti{\' e}rrez \cite{lozano2018complex} and information obtained from the LBSN data extracted in this work. The results indicate that, for most countries, there is a high correlation between data from LBSNs, specifically Foursquare, and official data. For these countries, LBSN data can satisfactorily represent the reality of the global tourism network.

The remainder of this document is presented as follows. Section~\ref{sec:relatedwork} briefly discusses the main related works and how this article differs from other previously published. Section~\ref{sec:methodology}  introduces methodological aspects of this study. Section~\ref{sec:results} presents and discusses the obtained results. Finally, the conclusions and future work are presented in Section~\ref{sec:conclusions}.

\section{Related Work} \label{sec:relatedwork}

In this section, we present studies that explore tourist mobility with traditional and social networks data. For instance, Migu{\' e}ns and Mendes \cite{miguens2008travel} analyze the distributions of degree and strength in a directed international mobility graph where nodes are countries, and the weight of the edges represents the number of tourists traveling between those countries, generated based on data from a 2004 WTO survey. They discovered that degree distribution is random, but the weighted network is scale-free (follows a power-law distribution), demonstrating the importance of considering the weight of edges in a tourism graph. In addition, they investigated the correlation between degree and strength, finding that, for the analyzed network, In-degree has a strong correlation with In-strength. At the same time, Out-strength grows almost quadratically concerning Out-degree. Zheng et al. \cite{zheng2009mining} analyzed 107 GPS logs of users over one year. The authors concluded that the mobility of residents and tourists is different; in addition, tourists' behavior is influenced by their relationships and traveling experiences. Mobility is an under-explored behavior in part because it is hard to obtain an appropriate description for that \cite{lew2006modeling, fennell1996tourist}.

There are also several works using social media data to study the mobility of tourists. For example, Hawelka et al. \cite{hawelka2014geo} use georeferenced Twitter data to study international mobility. When analyzing the network topology between nations, the emerging clusters respect real geopolitical boundaries, even if physical distance is not used as a variable for the group. To validate the results, they were contrasted  with statistics from the international economic forum, specifically on international tourist arrivals and tourism revenue, observing a strong linear correlation. Senefonte et al. \cite{senefonteSocInfo2020} study regional influences on the behavior of tourists and residents in the context of mobility using Foursquare data. They show, among other results, that the ability to explore the cultural characteristics of each nationality in different destinations leads a promising way to improve recommendation systems for points of interest and other services to particular groups of tourists. Ferreira et al. \cite{FERREIRA2020240} show that it is possible to go one step forward in the understanding of tourists' mobility, identifying where and when places are more important to users in different cities, by using social media data.


Belyi et al. \cite{belyi2017global} argue that human mobility is very complex because of its variety, including very different patterns such as tourist visits without recurrence or permanent migrations. Furthermore, each data source may emphasize a specific aspect of mobility, showing a distinct bias. Their study combines three data sources: Flickr (representing leisure activities and sightseeing), Twitter (any activity in an environment with internet access, whether business or leisure), and official United Nations migration data. This "multi-layer" model shows patterns that are not visible with the three layers separately, better fitting expectations when compared with other international relationship networks, such as common language and trade relation graphs. Comparisons are made in terms of networks' structure instead of directly comparing edges. The article also reports that country groups in the multi-layer network tend to be geographically connected, even though the physical distance between countries is not used in the algorithm. Another conclusion is that normalized weights of edges in the data sets extracted from Twitter and Flickr follow the same probability distribution with similar parameters. Still, the migration data tend to be more diverse.

Provenzano et al. \cite{provenzano2018mobility} analyze tourism within Europe using both WTO data and georeferenced Twitter data, concluding that there is a large overlap between the two data sets, even though social network data is more complete. Zhou et al.  \cite{zhou2016structure} study the international trade network by building a graph with edges only between a country and its largest economic partner and tries to justify the validity of this model using a triad census. As each country's trade is focused on a few other countries and several countries concentrate trade, it makes sense to focus on the most important relationship for each country.

The present work differs from similar efforts because its focus is on the extensive comparison between official data obtained using traditional ways, specifically those studied by Lozano and Guti{\' e}rrez \cite{lozano2018complex}, with mobility extracted using social media data. Although there are other comparisons between different data sources for the same context  \cite{provenzano2018mobility, belyi2017global}, they focus on specific cases, not providing a clear indication of how extensively social media data reflects characteristics obtained by official data regarding the global tourism network. This work is also different because it studies tourist mobility on a global scale, while many others focus on small regions, such as the study of Provenzano et al. \cite{provenzano2018mobility}, Leung et al. \cite{leung2012social}, and Piazzi et al. \cite{piazzi2011destinations}.
    
\section{Data and Methods}\label{sec:methodology}

This section presents the data used and the key methodological steps adopted in the present work.  We virtually replicate the main results obtained by Lozano and Guti{\' e}rrez \cite{lozano2018complex}. However, instead of using WTO survey, the work considers Foursquare data, whose details are presented in the next section. Additionally, and as one of the contributions of the present work, we perform at the end of the paper a multidimensional analysis joining all the metrics to show where -  in which countries - there are high correlations between datasets and where there are not.

\subsection{Data Description}

Two datasets are used in this work. The first one is composed of check-ins grouped by users obtained from the location-based social network Foursquare\footnote{https://foursquare.com.}. The same dataset was also used in \cite{FERREIRA2020240,senefonteSocInfo2020}. It represents globally shared data spanning four months, April–July of 2014. The total number of check-ins in the dataset is 19,716,435. The total number of countries in the dataset is 230, but only countries with more than 1000 check-ins are considered in this study, resulting in 117 countries. Appendix \ref{appendix1} shows all 117 countries considered in this study grouped by continent. Africa and Oceania are under-represented continents in the dataset due to: i) small population for Oceania; and ii) less access to the internet and weaker tourism industry for Africa.

Due to the origin of the dataset, it is important to consider the fact that it would be nearly impossible to acquire more recent data from the same source, as Foursquare is not as used as it was in 2014. However, it would be possible to acquire similar data from other social networks, such as Yelp. The fact the dataset spans three months does mean it also has a seasonal bias, as it would be autumn in the southern hemisphere and spring in the northern hemisphere. This may also be a cause of some of the disparities observed between the two datasets.

The other dataset contains official data from WTO surveys ("\textit{Tourism statistics‐ Arrivals of non‐resident visitors at national borders, by nationality}" and "\textit{Tourism statistics‐Outbound tourism‐ trips abroad by resident visitors to countries of destination (basis: arrivals in destination countries)}"). These data represent the number of tourists arriving in/leaving a country from/to every other country under study. The WTO dataset has been analyzed by Lozano and Guti{\' e}rrez \cite{lozano2018complex}; thus, the data and results are used in the present paper as a comparison basis with the Foursquare data. The WTO dataset under evaluation was collected originally in 2016.

\subsection{Graph Generation}

The global tourism network is modeled by graphs in this study. In the graphs obtained from Foursquare data, 117 different countries are considered. In the WTO data, there are 214 countries in the outbound tourism (Out) subgraphs and 148 countries in the inbound tourism (In). For 66 countries, there is no inbound data.

Similar to the assumptions made in \cite{senefonteSocInfo2020}, \cite{hawelka2014geo}, and \cite{FERREIRA2020240}, the country where a user of Foursquare made the most check-ins is considered the homeland of that user, and for all the remaining countries, such user is considered a tourist. For each user from a given country, all different countries visited are counted. A directed graph $G=(V,E)$ is then built, where the set $V$ of nodes represents the selected countries and a directed edge $e_{i,j} \in E$ with weight $w_{i,j} \in \mathbb{N}$ connects countries $v_i$ to $v_j \in V$ if the number $w_{i,j}$ of tourists who live in $v_i$ and have visited $v_j$ is greater than zero -- i.e., at least one check-in was made in $v_j$.

Following the proposal presented in the baseline work~\cite{lozano2018complex}, subgraphs $G_{in, k}$ (incoming tourists) and $G_{out, k}$ (outgoing tourists) are created according to the following definitions: $G_{in, k} = (V, E_{in, k})$ with $E_{in, k} \subset E$ and $e_{i,j} \in E_{in, k }$ if $e_{i,j}$ is among the $k \in \{1, 2, 3\}$ edges with the highest weight entering $v_j$; $G_{out, k} = (V, E_{out, k})$ with $E_{out, k} \subset E$ and $e_{i,j} \in E_{out, k}$ if $ e_{i,j}$ is among the $k \in \{1, 2, 3 \}$ edges with the highest weight coming out of $v_i$. Therefore, this study uses 6 subgraphs representing each of the two datasets. 

Different analyses are performed on the subgraphs, from basic statistics
-- obtained with the Python 3 packages NetworkX \cite{hagberg2008exploring} and Pandas \cite{reback2020pandas} -- to more complex structural properties. First, we replicated all analyses performed by Lozano and Guti{\' e}rrez \cite{lozano2018complex}, which are used as baselines for comparing with the results obtained from social media data.  Then we provide a final multidimensional analysis to show where there is correlation and where there is not.

\subsection{Country Rankings Based on Centrality Measures}

We use different centrality measures to 
rank the most important countries based on specific perspectives. First, we apply the PageRank centrality \cite{brin1998anatomy} in each subgraph. Next, we compute In-strength, Out-strength, and Betweenness centrality \cite{newman2018networks} for the Top-3 In and Top-3 Out subgraphs. We also compute In-degree for the Top-3 Out subgraph and Out-degree for the Top-3 In subgraph. These centralities are calculated using the NetworkX and Pandas libraries of the programming language Python. Further details on these centralities measures can be found in \cite{newman2018networks}.

\subsection{Hierarchical Clustering and Strongly Connected Components}\label{secMetHierar}

The strongly connected components of the Top-3 In and Top-3 Out subgraphs are determined for the Foursquare dataset by a NetworkX library function. By following the procedures applied in \cite{lozano2018complex}, components containing only one country are ignored.


A hierarchical clustering algorithm with \textit{average linkage} criteria~\cite{tan2016introduction} is used as in \cite{lozano2018complex}. Here it is accomplished in \textit{Python} with libraries \textit{Numpy} \cite{harris2020array}, \textit{NetworkX}, and \textit{Sklearn} \cite{scikit-learn}. In this analysis, it is necessary to represent graphs as distance matrices. This is done by normalizing the adjacency matrices of the graphs. Normalization occurs by row for the Top-3 Out subgraph and by column for the Top-3 In subgraph. This is done so that the sum of the normalized elements $n(w_{i,j })$ of the rows or columns of the matrix results in 1. Afterward, each element of the distance matrix is updated with the value $1-n(w_{i,j})$, transforming normalized affinity into the normalized distance. To make the comparison simpler, the dendrogram cut-off point is defined to generate the same number of clusters as defined by Lozano and Guti{\' e}rrez \cite{lozano2018complex}, excluding those with only one country.

\subsection{Cross-regional Tourism Flows}

In addition to the flow between countries, an analysis of cross-regional tourism flow is also carried out. The regions considered are the continents of North and South America, Europe, Africa, Asia, and Oceania. In the same way, as in the country analysis, directed graphs are used, now with notations corresponding to the continents: $G_{cont, in} = (V_{cont}, E_{cont})$ where $V_{cont}$ are the continents listed and the weight of $e_{i, j} \in E_{cont, in}$ is equal to the sum of the edge weights in the Top-3 In subgraph such that the country of origin is from the continent $ v_i$ and the destination country from the continent $v_j$. 

Transitions within the same continent, meaning tourists' travels to a different country in their home continent, were maintained, as in the reference study. For fairness in comparison with WTO data, the LBSN data are normalized by dividing the weight of each edge by the sum of the weight of all edges, thus obtaining the percentage of tourism flow that occurs between each pair of continents for each dataset.

\subsection{Motifs}

This section describes the techniques used to reveal the local patterns of the subgraphs: motif analysis, which is performed with the mfinder program \cite{milo2002network}. The triad census analysis, which has overlapping results with the motif ones, is shown in the Appendix \ref{secTriadCensus}. 

 Motif analysis consists in counting and classifying the patterns of edges between each set of nodes in the graph. A comparison with random networks of the input graph size is made to determine the most relevant patterns. 

The algorithm of mfinder uses the Z-value of the frequency of a particular motif in relation to the frequency of that motif in a set of random networks of the same size to choose the most relevant motifs in the input graph. Taking advantage of this, the results of the motif census using three nodes for Foursquare and WTO data have been compared through the percentage difference of the Z-values for each motif. This metric is used because it diminishes the influence of the subgraph size difference for the two datasets.

\section{Results} \label{sec:results}

This section presents the results obtained
using the Foursquare dataset; they are further compared with those obtained by~\cite{lozano2018complex} using the WTO dataset. Full details on the results for the WTO dataset may be consulted in~\cite{lozano2018complex}.

\subsection{Overview of Subgraphs}
\label{sec:subgraph}

\begin{figure*}[htttb!]
\centering
\subfigure[Top-1 In]
{\includegraphics[width=.31\textwidth]{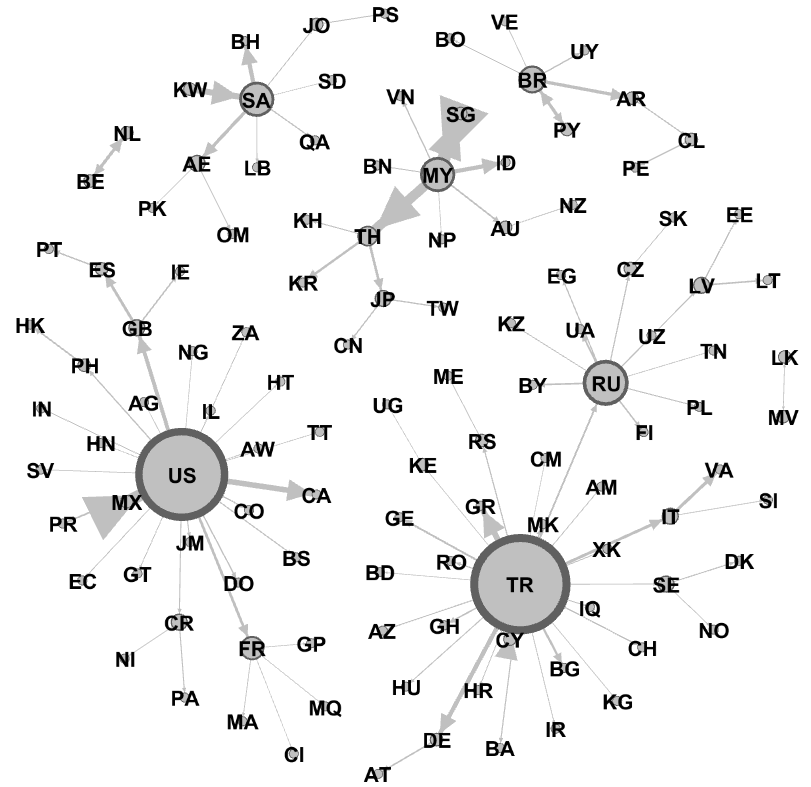}\label{grafo_in_top1}}\hspace{0.5cm}
  \subfigure[Top-2 In]
{\includegraphics[width=.31\textwidth]{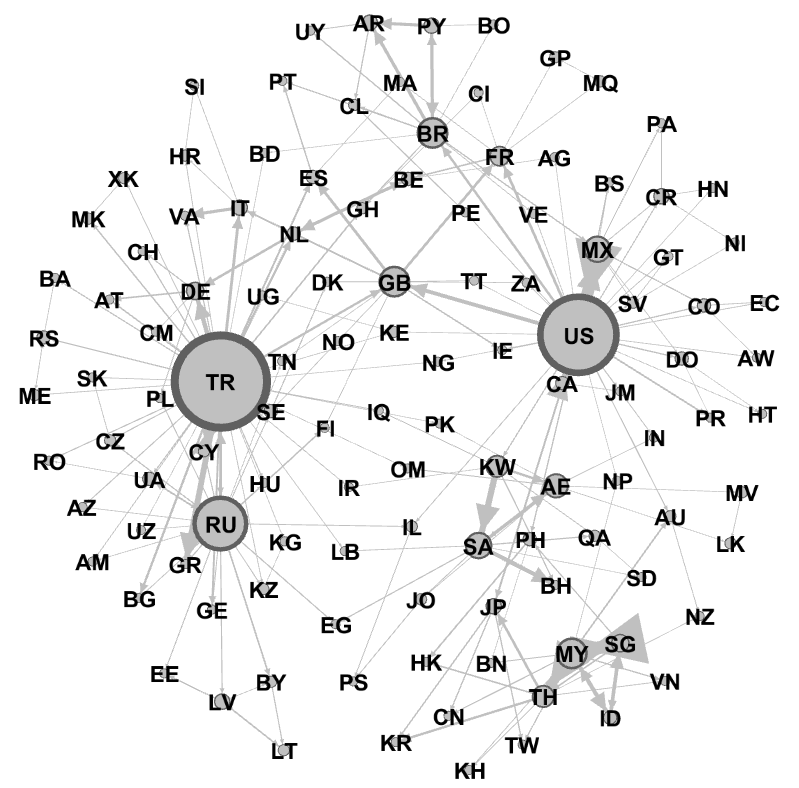}\label{grafo_in_top2}}\hspace{0.5cm}
 \subfigure[Top-3 In]
{\includegraphics[width=.31\textwidth]{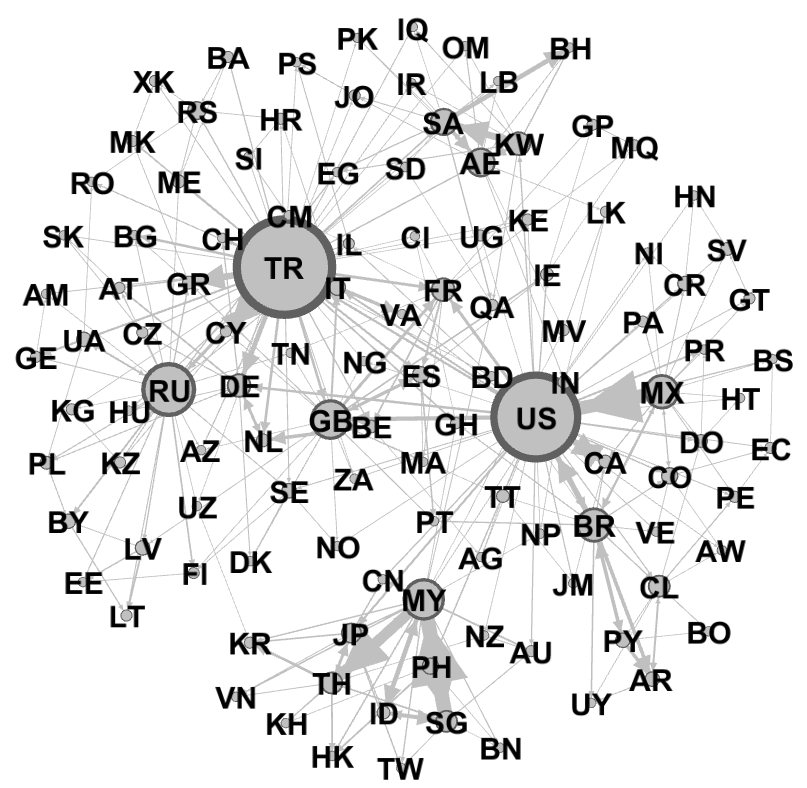}\label{grafo_in_top3}}
 \caption{Top-k In graphs generated for the Foursquare dataset. Nodes represent countries identified by their ISO-3166 alpha-2 country code. The size of nodes is proportional to their Out-degree, and the size of edges to their weight.}
\label{figNetworksIn}
\end{figure*}

\begin{figure*}[htttb!]
\centering
\subfigure[Top-1 Out]
{\includegraphics[width=.31\textwidth]{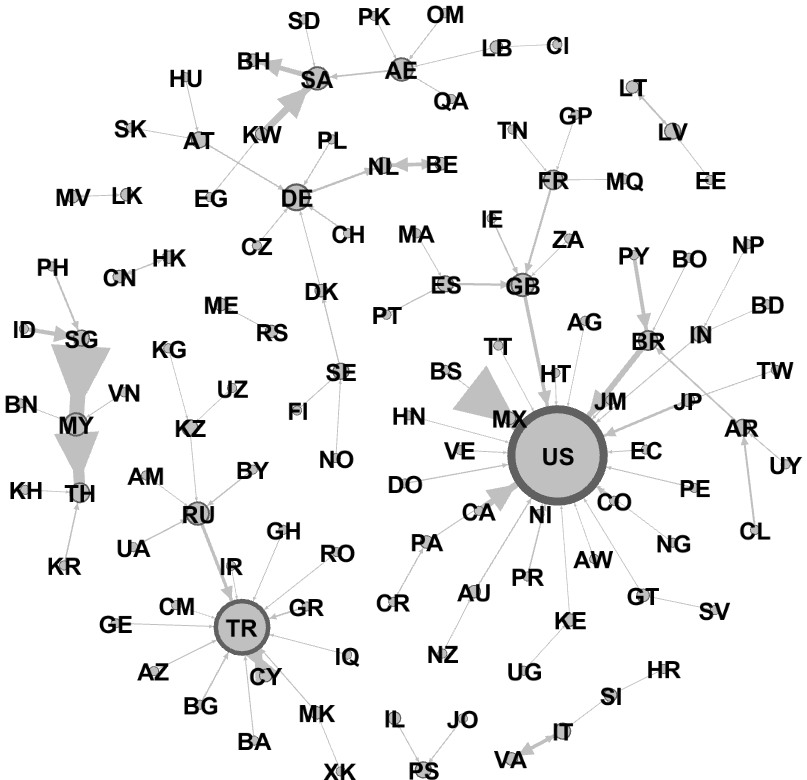}\label{grafo_out_top1}}\hspace{0.5cm}
  \subfigure[Top-2 Out]
{\includegraphics[width=.31\textwidth]{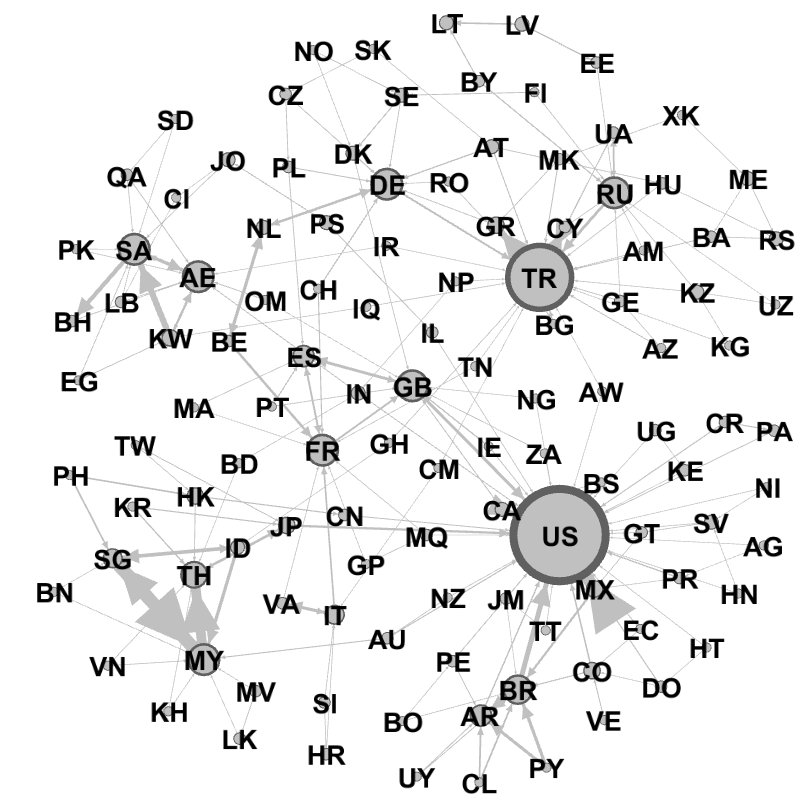}\label{grafo_out_top2}}\hspace{0.5cm}
 \subfigure[Top-3 Out]
{\includegraphics[width=.31\textwidth]{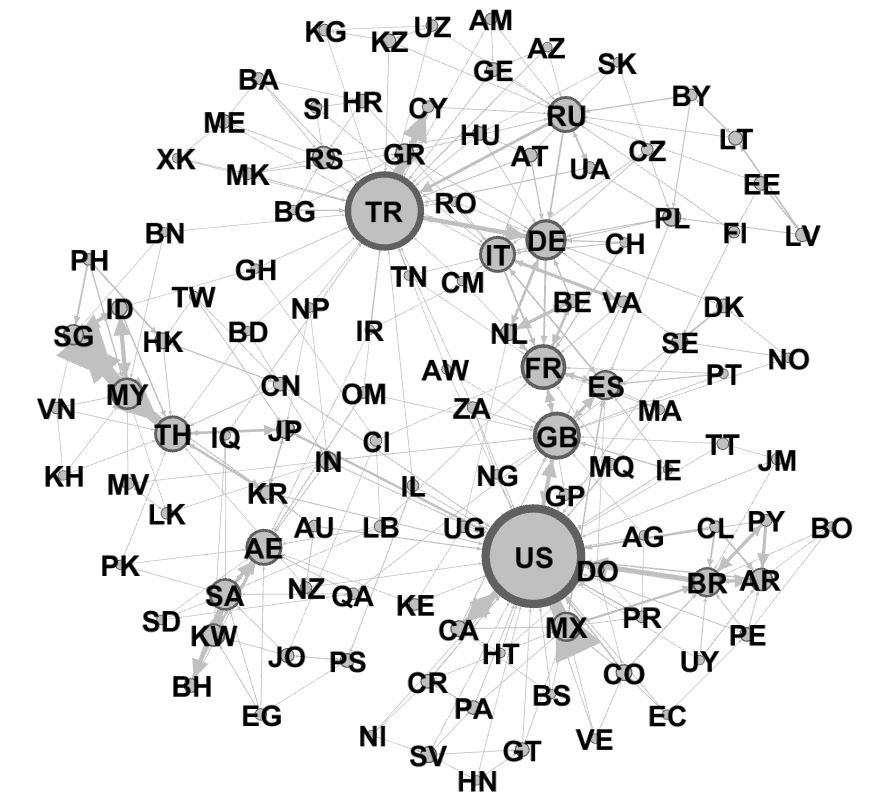}\label{grafo_out_top3}}
 \caption{Top-k Out graphs generated for the Foursquare dataset. Nodes represent countries identified by their ISO-3166 alpha-2 country code. The size of nodes is proportional to their In-degree, and the size of edges to their weight.}
\label{figNetworksOut}
\end{figure*}

First, subgraphs created with Foursquare data are evaluated. Figures \ref{grafo_in_top1}, \ref{grafo_in_top2}, and \ref{grafo_in_top3} show the subgraphs for the Top-1, Top-2, and Top-3 In mobility graphs, respectively, whereas, Figures \ref{grafo_out_top1}, \ref{grafo_out_top2}, and \ref{grafo_out_top3} show the Top-1, Top-2, and Top-3 Out mobility graphs, respectively. Each country in the network is represented by a node identified by the country's ISO-3166 alpha-2 country code. The thickness of edges is proportional to their weight, and the size of the nodes is proportional to their Out-degree for Top-k In subgraphs and their In-degree for Top-k Out subgraphs. Note that the United States (US) and Turkey (TR) have the greatest importance in the network. This phenomenon is intuitive for the US, as a great number of American tourists is expected in other countries as well as a great number of foreign tourists is expected in the US. On the other hand, the importance of Turkey is due to the great popularity of the social network in that country during the studied period. In addition, it is possible to identify regional tourism centers by analyzing the subgraphs. In almost all of them, Russia (RU), Malaysia (MY), Brazil (BR), Saudi Arabia (SA), Mexico (MX), Germany (DE), France (FR), Italy (IT), and the United Kingdom (GB) have a secondary position.

From these observations, it is possible to conclude that the United States is the largest recorded source and destination of tourists in the network. The fact that countries such as the United Kingdom, France, and Germany stand out more in the Top-k Out graphs are indications that those countries are more important globally as destinations than as tourist sources.

Comparing these graphs with those generated from WTO data, the most striking difference is the outstanding representation of Turkey in the dataset. This finding is important to keep in mind, especially for other analyses that are presented below. Different values of $k$ provide particular information. For $k=1$, the Top-1 In/Out subgraphs are extremely focused, useful for identifying the most critical relationship of each country. The Top-2 and Top-3 In/Out subgraphs, i.e., $k=2$ and $k=3$, are useful to identify small communities and relationships between more than two countries. This type of relationship can be helpful, for instance, to recommend countries to visit.

\subsection{General Statistics}\label{sec:quantities}

Tables~\ref{grandezas_grafo_out} and~\ref{grandezas_grafo_in} summarize general statistics about the created subgraphs. It is interesting to note that some information in the tables is determined by how the subgraphs are constructed: the number of edges for the Top-k In and Out subgraphs is determined by multiplying the number of countries by k. The average degree is equal to k, and the graph density is equal to k divided by the number of countries minus one, which means the subgraphs tend to have very low densities, as the number of countries is far greater than k.  It is also notable that the transitivity of Top-1 graphs is always 0, as it is impossible to form triangles in these subgraphs. In the following statistics, only average strength takes into account the weight of the edges.

\begin{table}[htbp]
\caption{Statistics of the Top-k Out subgraphs.}
\scriptsize
\begin{center}
\begin{tabular*}{\tblwidth}{@{}LLLL@{}}
\toprule
\textbf{Quantity}&\multicolumn{3}{c}{\textbf{Subgraph}} \\
& \textbf{\textit{Top-1 Out}}& \textbf{\textit{Top-2 Out}}& \textbf{\textit{Top-3 Out}} \\
\midrule
Number of edges & 117 & 234 & 351  \\
Density & 0.009 & 0.017 & 0.026  \\
Average geodesic distance & 1.877 & 3.81 & 3.888  \\
Diameter & 5 & 13 & 12  \\
Average degree & 1 & 2 & 3  \\
In-degree centralization & 0.210 & 0.289 & 0.342 \\
average strength & 208,085 & 316,632 & 388,513  \\
number of  & 11/95/ & 29/176/ & 51/249/  \\
mutual/assymetric/null dyads & 6680 & 6581 & 6486  \\
arc reciprocity & 0.188 & 0.248 & 0.291  \\
transitivity & 0.000 & 0.410 & 0.413  \\
\bottomrule
\end{tabular*}
\label{grandezas_grafo_out}
\end{center}
\end{table}

\begin{table}[htbp]
\caption{Statistics of the Top-k In subgraphs.}
\scriptsize
\begin{center}
\begin{tabular*}{\tblwidth}{@{}LLLL@{}}
\toprule
\textbf{Quantity}&\multicolumn{3}{c}{\textbf{Subgraph}} \\
& \textbf{\textit{Top-1 In}}& \textbf{\textit{Top-2 In}}& \textbf{\textit{Top-3 In}} \\
\midrule
Edge number & 117 & 234 & 351  \\
Density & 0.009 & 0.017 & 0.026  \\
Average geodesic distance & 1.769 & 2.729 & 3.129  \\
Diameter & 4 & 9 & 9  \\
Average degree & 1 & 2 & 3  \\
Out-degree centralization & 0.210 & 0.342 & 0.430  \\
Average strength & 265,932 & 367,726 & 437,726  \\
number of & 7/103/ & 18/198/ & 35/281/  \\
mutual/assymetric/null dyads & 6676 & 6570 & 6470\\
Arc reciprocity  & 0.120 & 0.154 & 0.199  \\
Transitivity & 0.000 & 0.029 & 0.049  \\
\bottomrule
\end{tabular*}
\label{grandezas_grafo_in}
\end{center}
\end{table}

The densities of the diad types are not comparable between WTO and Foursquare data, as the density of the subgraphs directly influences the densities of the diads. Analyzing the average strength, the number of tourists represented by WTO data is around ten times greater than the Foursquare data for the inbound (In) subgraphs and from three to five times greater for the outbound (Out) subgraphs.

Average geodesic densities between Foursquare and WTO data 
are mostly similar. In-degree centralization is smaller in Foursquare data than in WTO data, but Out-degree centralization is greater in Foursquare than in WTO data. Arc reciprocity is slightly greater in the Top-k Out Foursquare subgraphs than in the WTO ones, but between Top-k In subgraphs, it stays basically the same.

\subsection{Centrality Analyses}
\label{sec:centralid}

We first study the results from the PageRank centrality. Table~\ref{pagerank} shows the 20 countries with the highest PageRank centralities. Countries highlighted in bold are among the top 20 countries of all three Top-k In subgraphs or all three Top-k Out subgraphs. Countries in bold with an asterisk following their names appear in all six rankings. 

\begin{table}[htbp]
\caption{Countries with the Top-20 highest Pagerank centralities for each Foursquare subgraph. Countries in bold appear in all 3 Top-k In or all Top-k out rankings. Countries in bold with an asterisk following their name appear in all six rankings.}
\scriptsize
\begin{center}
\begin{tabular*}{\tblwidth}{@{}LLL@{}}
\toprule
\multicolumn{3}{c}{\textbf{Subgraph}} \\
\textbf{\textit{Top-1 Out}}& \textbf{\textit{Top-2 Out}}& \textbf{\textit{Top-3 Out}} \\
\midrule
\textbf{US*}&\textbf{US*}&\textbf{US*} \\
\textbf{Mexico} & \textbf{Turkey} & \textbf{United Kingdom} \\
\textbf{Turkey} & \textbf{Mexico} & France \\
\textbf{Cyprus} & Canada & \textbf{Turkey} \\
\textbf{Netherlands*} & \textbf{Cyprus} &  Spain\\
\textbf{Saudi Arabia} & \textbf{Saudi Arabia} & \textbf{Mexico} \\
Belgium & Greece & Canada \\
\textbf{Bahrain} & UAE & \textbf{Germany} \\
\textbf{Malaysia*} & \textbf{United Kingdom} & \textbf{Saudi Arabia} \\
\textbf{Thailand} & \textbf{Malaysia*} & Italy \\
Italy & Spain & UAE \\
Vatican & \textbf{Germany} & Belgium \\
Latvia & \textbf{Bahrain} & \textbf{Netherlands*} \\
Palestine & France & \textbf{Malaysia*} \\
Israel & \textbf{Thailand} & \textbf{cyprus} \\
Lithuania & Brazil & Greece \\
\textbf{Germany} & \textbf{Netherlands*} & \textbf{Thailand} \\
\textbf{United Kingdom} & Singapore & Vatican  \\
Sri Lanka & Kuwait & Kuwait \\
Maldives & Japan & \textbf{Bahrain} \\
\bottomrule
\toprule
\multicolumn{3}{c}{\textbf{Subgraph}} \\
\textbf{\textit{Top-1 In}}& \textbf{\textit{Top-2 In}}& \textbf{\textit{Top-3 In}} \\
\midrule
\textbf{Maldives} & Guadeloupe & Martinique \\
\textbf{Netherlands*} & Martinique  & Guadeloupe  \\
\textbf{Belgium} & \textbf{Sri Lanka} & Macedonia \\
\textbf{Sri Lanka} & \textbf{Maldives} & Kosovo \\
Peru & Macedonia & \textbf{Sri Lanka} \\
Chile & Kosovo & \textbf{Maldives} \\
Portugal & Latvia & Kazakhistan \\
\textbf{Malaysia*} & Lithuania & Quirguistan \\
Brazil & \textbf{Netherlands*} & Quenia \\
\textbf{US*} & \textbf{Belgium} & Uganda \\
Turkey & \textbf{US*} & \textbf{Netherlands*} \\
Oman & Serbia & \textbf{Malaysia*} \\
Saudi Arabia & \textbf{Malaysia*} & \textbf{US*} \\
Austria & Montenegro & Serbia \\
Argentina & Germany & Costa Rica \\
Slovakia & Palestine & Latvia \\
New Zealand & Singapore & \textbf{Belgium} \\
Vatican & Peru & El Salvador \\
Hong Kong & Estonia & Panama \\
Palestine & Saudi Arabia & Slovakia \\
\bottomrule
\end{tabular*}
\label{pagerank}
\end{center}
\end{table}

The countries' names written in bold in the Top-k Out subgraphs of the Foursquare base are the US, Mexico, Turkey, Cyprus, Netherlands, Saudi Arabia, Bahrain, Malaysia, Thailand, Germany, and the UK (11 out of 20 countries), and in the WTO dataset (obtained from \cite{lozano2018complex}), the US, Mexico, South Africa, Thailand, Botswana, Malaysia, France, Spain, Ukraine, Israel, Mauritius, Hong Kong, Benin, Greece, Ethiopia and the Philippines (16 out of 20 countries). Thus, it is clear that the classification of the centrality in the Top-k Out subgraphs is more homogeneous in WTO data than in Foursquare data. Furthermore, in the three subgraphs of the two datasets -- Top-1/2/3 Out --, the United States has the highest centrality. Turkey's disproportionate popularity in the Foursquare dataset has influenced its presence at the top of the three indexes as well as Cyprus, an important tourist destination for Turks. Finally, no African country appears in these indices, unlike the WTO data.

As shown in the second part of Table~\ref{pagerank}, for the Top-k In subgraphs of the Foursquare dataset, the countries in bold are Maldives, Netherlands, Belgium, Sri Lanka, Malaysia, and the US (6 of 20). The ones in the WTO dataset are the US, Canada, China, Hong Kong, Germany, Argentina, France, South Africa, the Russian Federation, Ukraine, and Uzbekistan (11 of 20). Interestingly, the countries with the highest centrality for the Foursquare Top-k In subgraphs are Maldives, Guadeloupe, and Martinique, with the US close to tenth place. At the same time, the US continues to dominate the Top-K In in the WTO dataset. This may indicate that, when considering individual countries, Foursquare data represents those that act primarily as tourist sources more accurately than those that serve as destinations; this last case is probably influenced by the LBSN bias.

Complementing the analysis of the studied countries, Tables~\ref{grandezas_paises_out} and \ref{grandezas_paises_in} show three centralities (In-strength, Out-strength, and Betweenness) common to the Top-3 In and Out subgraphs, whereas In-degree is shown only for Top-3 Out and Out-degree only for Top-3 In.


For simplicity, only the first 20 positions for each centrality measure are shown. Countries appearing in at least three columns are shown in bold for each table, whereas an asterisk identifies countries appearing in all four columns.

\begin{table}[htbp]
\scriptsize
\caption{In-degree, In-strength, Out-strength, and Betweenness centralities for the Top-3 Out Foursquare subgraphs. Countries in bold appear in the first 20 positions for at least three out of four centrality rankings shown in the table, and countries with asterisks appear in the first 20 positions for all the four centrality rankings.}

\begin{center}
\begin{tabular*}{\tblwidth}{@{}LLLLL@{}}
\toprule
Ranking & In-degree & In-strength & Out-strength & Betweenness  \\
\midrule
1 & \textbf{US*}&\textbf{US*}& \textbf{Malaysia} & \textbf{US*} \\
2 & \textbf{Turkey*} & \textbf{Malaysia} & \textbf{Singapore} & \textbf{United} \\
 & & & & \textbf{Kingdom*} \\
3 & \textbf{United} & \textbf{Thailand*} & \textbf{Turkey*} & \textbf{France} \\
&   \textbf{Kingdom*}  & & & \\
4 & \textbf{France} & \textbf{Turkey*} & \textbf{Mexico*} & \textbf{Turkey*} \\
5 & \textbf{Germany*} & \textbf{Singapore} & \textbf{US*} & \textbf{Germany*}\\
6 & \textbf{UAE} & \textbf{Germany*} & \textbf{Brazil*} & \textbf{Mexico*} \\
7 & \textbf{Russian} & \textbf{France} & Kuwait & Belgium \\
  & \textbf{Federation}& & &  \\
8 & \textbf{Thailand*} & \textbf{United} & \textbf{Saudi Arabia} & \textbf{Brazil*} \\
  & & \textbf{Kingdom*} & & \\
9 & \textbf{Italy} & \textbf{Saudi Arabia} & \textbf{United} & \textbf{Russian} \\
  & &  & \textbf{Kingdom*} & \textbf{Federation} \\
10 & \textbf{Malaysia} & \textbf{Brazil*} & Indonesia & \textbf{Spain*} \\
11 & \textbf{Saudi Arabia} & \textbf{Argentine} & \textbf{Thailand*} & \textbf{Italy} \\
12 & \textbf{Mexico*} & \textbf{Mexico*} & Canada & \textbf{UAE} \\
13 & \textbf{Spain*} & \textbf{Spain*} & Cyprus & Oman \\
14 & \textbf{Brazil*} & \textbf{UAE} & Paraguay & \textbf{Thailand*}\\
15 & Kuwait & Greece & Belgium & Greece \\
16 & \textbf{Argentine} & Cyprus & Netherlands & Cyprus \\
17 & Serbia & Indonesia & \textbf{Italy} & Ukraine \\
18 & \textbf{Singapore} & Canada & \textbf{Russian} & \textbf{Argentine}  \\
   & &  & \textbf{Federation} & \\
19 & Japan & Bahrain & \textbf{Germany*} & Japan  \\
20 & Poland & Netherlands & \textbf{Spain*} & India \\
\bottomrule
\end{tabular*}
\label{grandezas_paises_out}
\end{center}
\end{table}

\begin{table}[htbp]
\scriptsize
\caption{Out-degree, Out-strength, In-strength, and Betweenness centralities for the Top-3 In Foursquare subgraphs. Countries in bold appear in the first 20 positions for at least three out of four centrality rankings shown in the table, and countries with asterisks appear in the first 20 positions for all the four centrality rankings.}
\begin{center}
\begin{tabular*}{\tblwidth}{@{}LLLLL@{}}
\toprule
Ranking & Out-degree & Out-strength & In-strength & Betweenness  \\
\midrule
1 & \textbf{Turkey}&*\textbf{Turkey}*& \textbf{US*} & \textbf{Turkey*} \\
2 & \textbf{US*} & \textbf{US*} & \textbf{Malaysia*} & \textbf{US*} \\
3 & \textbf{Russian} & \textbf{Malaysia*} & \textbf{Singapore} & \textbf{Russian} \\
  & \textbf{Federation} & & & \textbf{Federation}\\
4 & \textbf{Malaysia*} & \textbf{Singapore} & \textbf{Thailand} & \textbf{Brazil*} \\
5 & \textbf{United} & \textbf{Mexico*} & \textbf{Turkey*} & Belarus\\
  & \textbf{Kingdom*} & & & \\
6 & \textbf{Mexico*} & \textbf{Brazil*} & \textbf{Germany*} & \textbf{United*} \\
  & & & & \textbf{Kingdom*} \\
7 & \textbf{Brazil*} & \textbf{Russian} & \textbf{Singapore} & Paraguay \\
  & & \textbf{Federation} & & \\
8 & \textbf{UAE*} & \textbf{Singapore} & \textbf{Mexico*} & Poland \\
9 & \textbf{Singapore} & \textbf{Kuwait} & Argentine & \textbf{Germany*} \\
10 & \textbf{Kuwait} & \textbf{United} & \textbf{Cyprus} & Ukraine \\
  & & \textbf{Kingdom*} & & \\
11 & \textbf{France*} & \textbf{Thailand} & \textbf{United} & \textbf{Mexico*} \\
  & & & \textbf{Kingdom*} & \\
12 & \textbf{Germany*} & Indonesia & Greece & \textbf{Cyprus}\\
13 & \textbf{Singapore} & \textbf{Cyprus} & \textbf{France*} & Latvia \\
14 & \textbf{Thailand} & \textbf{Germany*} & \textbf{Brazil*} & \textbf{UAE*}\\
15 & Chile & \textbf{Netherlands} & Canada & \textbf{Netherlands} \\
16 & \textbf{Japan} & \textbf{Japan} & \textbf{Netherlands} & \textbf{Spain} \\
17 & Serbia & Italy & Indonesia & \textbf{Malaysia*} \\
18 & Colombia & Belgium & \textbf{UAE*} & \textbf{France*}  \\
19 & Philippines & \textbf{France*} & \textbf{Spain} & \textbf{Kuwait}  \\
20 & \textbf{Spain} & \textbf{UAE*} & Italy & \textbf{Japan} \\
\bottomrule
\end{tabular*}
\label{grandezas_paises_in}
\end{center}
\end{table}

One common characteristic of the centrality rankings is the importance of the United States in the Top-3 Out rankings (Table \ref{grandezas_paises_out}),  even though it does not take the first position in the Foursquare Out-strength ranking. The fact that the United States is less dominant in the output rankings (Table \ref{grandezas_paises_in}), appearing at the second place in the WTO centrality rankings and the first place only in one of the Foursquare rankings (In-strength),  indicates that the United States plays a more predominant role as a "source" than as a destination in both datasets.  Also,  Turkey is, once again, disproportionately present in the Foursquare dataset, being first in both Out-degree and Out-strength rankings for the Top-3 In Foursquare subgraphs, even though it does not appear in the Top-20 of the equivalent WTO rankings. To provide fairness in the comparison, Tables \ref{comp_grandezas_paises_out} and~\ref{comp_grandezas_paises_in} show the rankings of countries in the WTO datasets and, in parentheses, the position of that country in the corresponding Foursquare ranking. Cells marked with "NA" indicate countries with less than 1000 Foursquare check-ins registered in their territory, ignored in the graph.

\begin{table}[htbp]
\scriptsize
\caption{Comparison of the country rankings in the Top-3 Out subgraph. Countries are arranged according to their position in the WTO rankings, and the number in parentheses is their position in the Foursquare ranking. If a country does not appear in the Foursquare ranking, the position is replaced by "NA."}
\begin{center}
\begin{tabular*}{\tblwidth}{@{}LLLLL@{}}
\toprule
WTO & In-degree  & In-strength & Out-strength & Betweenness  \\
Ranking & & & & \\ 
\midrule
1 & US(1)  & US(1) & US(5) & US(1) \\
2 & Malaysia(11) & Spain(13) & Germany(19) & France(3) \\
3 & South Africa(77) & France(7) & Canada(12) & Greece(15) \\
4 & Canada(27) & Ukraine(32) & China(38) & Cyprus(16) \\
5 & Ukraine(76) & Thailand(3) & United & Spain(10) \\
& & & Kingdom(9) & \\
6 & Thailand(8) & Malaysia(2) & Singapore(2) & Malaysia(26) \\
7 & Greece(22) & Hong Kong(29) & Mexico(4) & Ukraine(17) \\
8 & Spain(13) & Mexico(12) & Russian & Andorra(NA) \\
& & & Federation(18) & \\
9 & Benim(NA) & Canada(18) & France(21) & Philippines(64) \\
10 & France(4) & Greece(15) & Italy(17) & Thailand(14) \\
11 & Israel(66) & South Africa(79) & Netherlands(16) & South Africa(54) \\
12 & Brazil(14) & Ireland(96) & Switzerland(31) & Canada(49) \\
13 & Colombia(24) & Indonesia(17) & Spain(20) & Sri Lanka(110) \\
14 & Angola(NA) & Brazil(10) & Japan(22) & Mexico(6) \\
15 & Ethiopia(NA) & Uzbekistan(82) & Moldavia(NA) & Brazil(8) \\
16 & Mali(NA) & Andorra(NA) & Malaysia(1) & Mauritius(NA) \\
17 & Peru(37) & Peru(51) & South Korea(29) & Guadeloupe(100) \\
18 & Barbados(NA) & Cambodia(87) & Indonesia(10) & Dominica(NA) \\
19 & Botswana(NA) & Philippines(42) & Belarus(30) & Hong Kong(30) \\
20 & Mauritius(NA) & Botswana(NA) & Portugal(46) & Antigua \\
& & & & and Barbuda(87)\\
\bottomrule
\end{tabular*}
\label{comp_grandezas_paises_out}
\end{center}
\end{table}

\begin{table}[htbp]
\scriptsize
\caption{Comparison of the country rankings in the Top-3 In subgraph. Countries are arranged according to their position in the WTO rankings, and the number in parentheses is their position in the Foursquare ranking. If a country does not appear in the Foursquare ranking, the position is replaced by "NA."}
\begin{center}
\begin{tabular*}{\tblwidth}{@{}LLLLL@{}}
\toprule
WTO & Out-degree & Out-strength & In-strength & betweenness \\  
Ranking & & & & \\
\midrule
1 & US(2) & Hong  & China (32)& US(2) \\
  & & Kong(103) & & \\
2 & United  & China(76) & Hong & China (69) \\
  & Kingdom(5) &  &  Kong(26) & China (69) \\
3 & France(11) & US(2) & US(1) & Canada(54) \\
4 & Canada(36) & Germany(16) & Italy(20) & Russian \\
  & & & & Federation(3) \\
5 & China (76)& United & Spain(19) & United \\
  & & Kingdom(10) & & Kingdom (6)\\
6 & Germany (12)& France(21) & Poland(52) & Ukraine(10) \\
7 & Russian & Canada(14) & France(13) & Macao(NA)  \\
  & Federation(3) & & & \\
8 & Japan(16) & Russian & Macao(NA) & Romania(97) \\
  & & Federation(7) & & \\
9 & Australia(59) & Macao(NA) & Mexico(8) & Taiwan (96)\\
10 & New  & Singapore(4) & Ukraine(35) & Indonesia(35) \\
   & Zealand(116) & &  & \\
11 & Indonesia(37) & Mexico(5) & Malaysia(2) & Bulgaria(117) \\
12 & South & Ukraine(29) & Russian & Japan(20) \\
   & Africa(109) & & Federation(29) & \\
13 & Argentine(22) & Czech & Canada(15) & France(18) \\
13 &  & Republic(43) & &  \\
14 & Brazil(7) & Switzerland(72) & Sweden(58) & Egypt(70) \\
15 & Italy(21) & Taiwan(99) & Turkey(5) & Greece(31) \\
16 & South  & Saudi Arabia(8) & United & Tanzania(NA) \\
   & Korea(63) & & Kingdom(11) & \\
17 & Netherlands(39) & Malaysia(3) & Thailand(4) & Saudi \\
   &  &  &  & Arabia(22) \\
18 & Saudi Arabia(9) & Slovakia(90) & South Africa(88) & Turkey(1) \\
19 & Taiwan(99) & Kazakhstan(61) & South & Australia(55) \\
19 & Taiwan(99) & Kazakhstan(61) & Korea(25) & Australia(55) \\
20 & Malaysia(4) & Moldavia(NA) & Singapore(3) & Malaysia(17) \\
\bottomrule
\end{tabular*}
\label{comp_grandezas_paises_in}
\end{center}
\end{table}

Note that some countries are always in similar positions regardless of the dataset, such as the United States and Thailand in the Top-3 Out rankings. A few others appear in very different positions, such as Israel (for In-degree ranking it appears at 11\textsuperscript{th} in the WTO Top-3 Out  and 66\textsuperscript{th} in Foursquare) and Ireland (for In-strength ranking it appears at 12\textsuperscript{th} in the WTO Top-3 Out and 96\textsuperscript{th} in Foursquare). This is probably caused by the smaller number of check-ins in the dataset made by inhabitants of these smaller countries, leading to a less accurate model in these cases.

\subsection{Strongly Connected Components}
\label{sec:components}

An analysis of the strongly connected components in the Top-3 In/Out subgraphs is also performed for both datasets. Full lists of the countries in each component together with their geographical displacement are shown in Appendix~\ref{appendix3}.

\begin{figure}[htbp]
\centerline{\includegraphics[width=.48\textwidth]{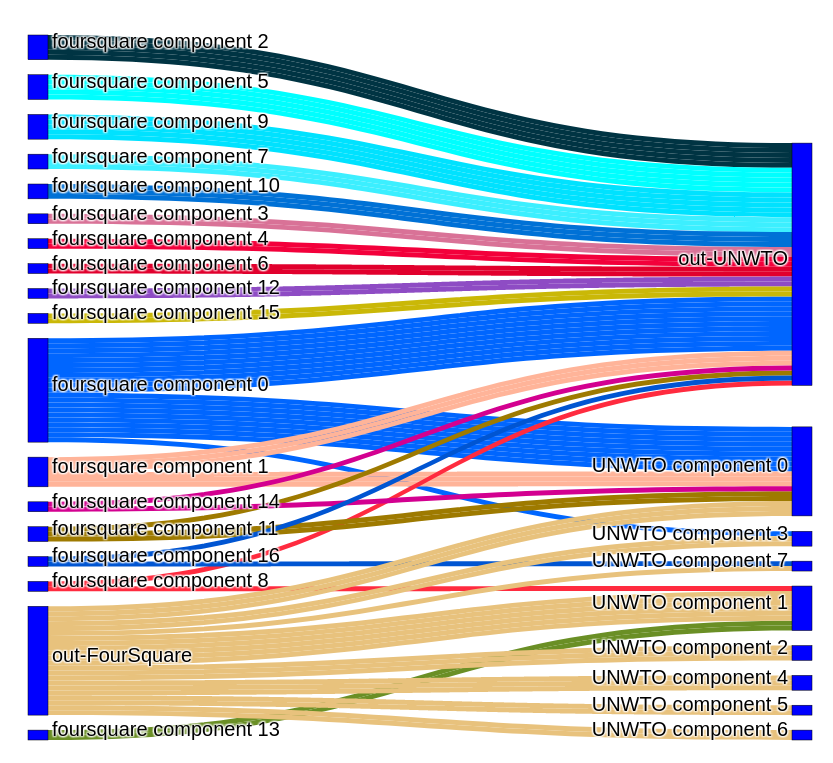}}
\caption{Parallel set representation of the strongly connected components of the Top-3 Out subgraphs for both datasets. On the left, we have Foursquare components and on the right, WTO components. The colored curves join the positions of each country in each datasets' components.}
\label{conjuntos_paralelos_out}
\end{figure}

Figures~\ref{conjuntos_paralelos_in} and \ref{conjuntos_paralelos_out} help us to understand the results. They illustrate the correspondence between Foursquare and WTO regarding this metric, where on the left side are the connected components for the Foursquare data, and on the right are those of the WTO data. For the Top-3 Out subgraph, there is a large component in the WTO data that is made up of many members from two of the Foursquare components. In both datasets' Top-3 In subgraphs, we have very similar small components strongly tied to geographic regions: South American, South Asian, Central American, Eastern European, and Arabic components. Note that in both Top-3 In and Out cases, we observe the tendency of smaller components linked to the same bigger one in the other dataset. This indicates considerable similarities among the results despite particular characteristics of the datasets that could have induced a component bigger or smaller.

\begin{figure}[htbp]
\centerline{\includegraphics[width=.49\textwidth]{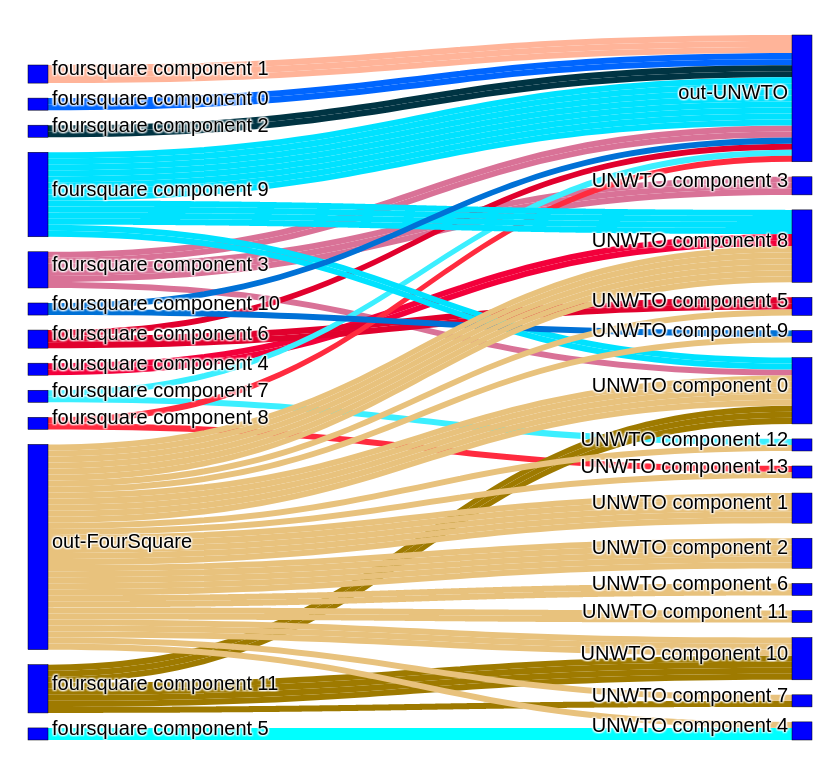}}
\caption{Parallel set representation of the strongly connected components of the Top-3 In subgraphs for both datasets. On the left, we have Foursquare components, and on the right, WTO components. The colored curves join the positions of each country in each datasets' components.}
\label{conjuntos_paralelos_in}
\end{figure}

\subsection{Hierarchical Clustering}

\begin{figure*}[httt!]
\centering
\subfigure[Foursquare]
{\includegraphics[width=.48\textwidth]{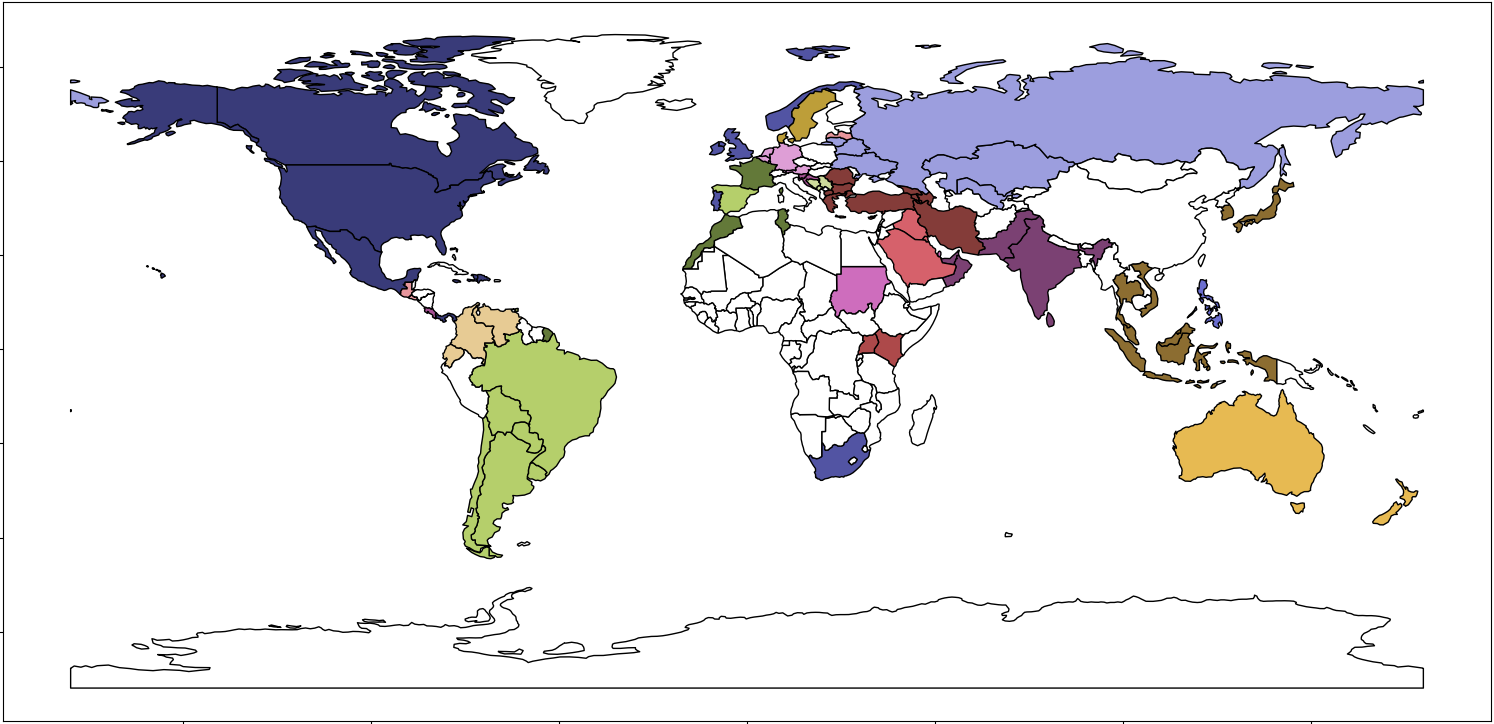}\label{fig_clusters_in}}\hspace{0.5cm}
  \subfigure[WTO]
{\includegraphics[width=.48\textwidth]{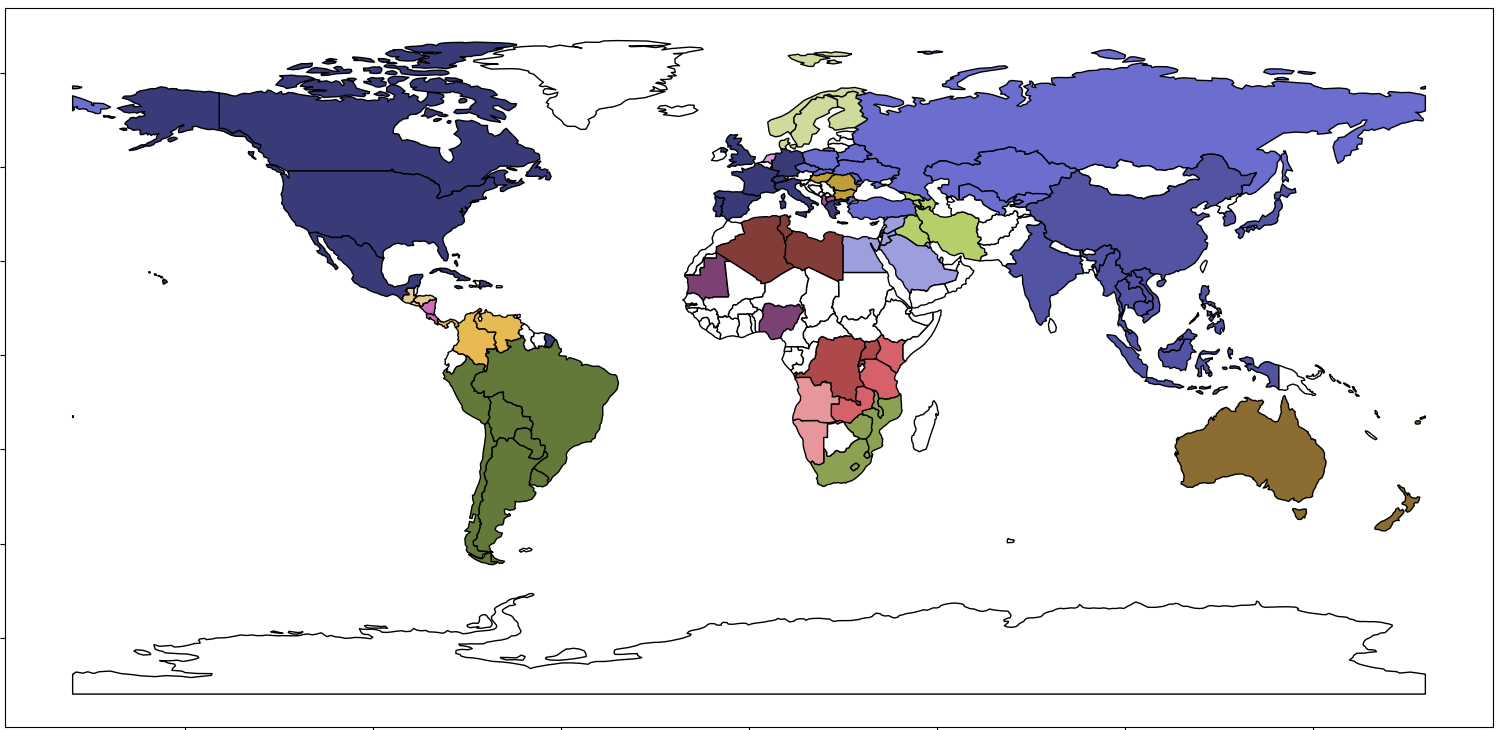}\label{fig_clusters_in_wto}}
 \caption{Clusters found for the Top-3 In graph generated for the Foursquare and WTO datasets. Colors in each figure represent a specific cluster. Countries in white have not been included in the model or have been ignored because they are part of a single-country component/cluster.}
\label{figTop3InComparison}
\end{figure*}

\begin{figure*}[httt!]
\centering
\subfigure[Foursquare]
{\includegraphics[width=.48\textwidth]{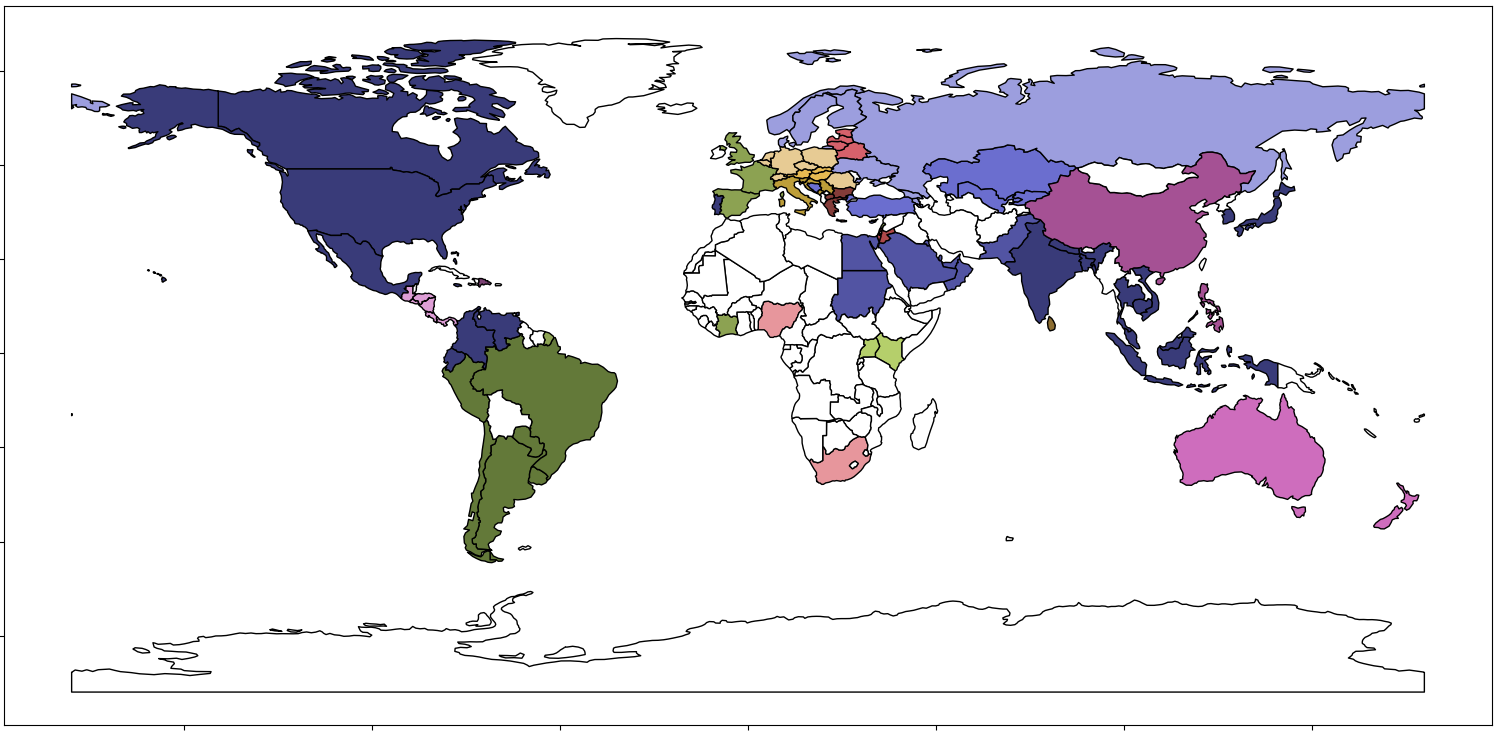}\label{fig_clusters_out}}\hspace{0.5cm}
  \subfigure[WTO]
{\includegraphics[width=.48\textwidth]{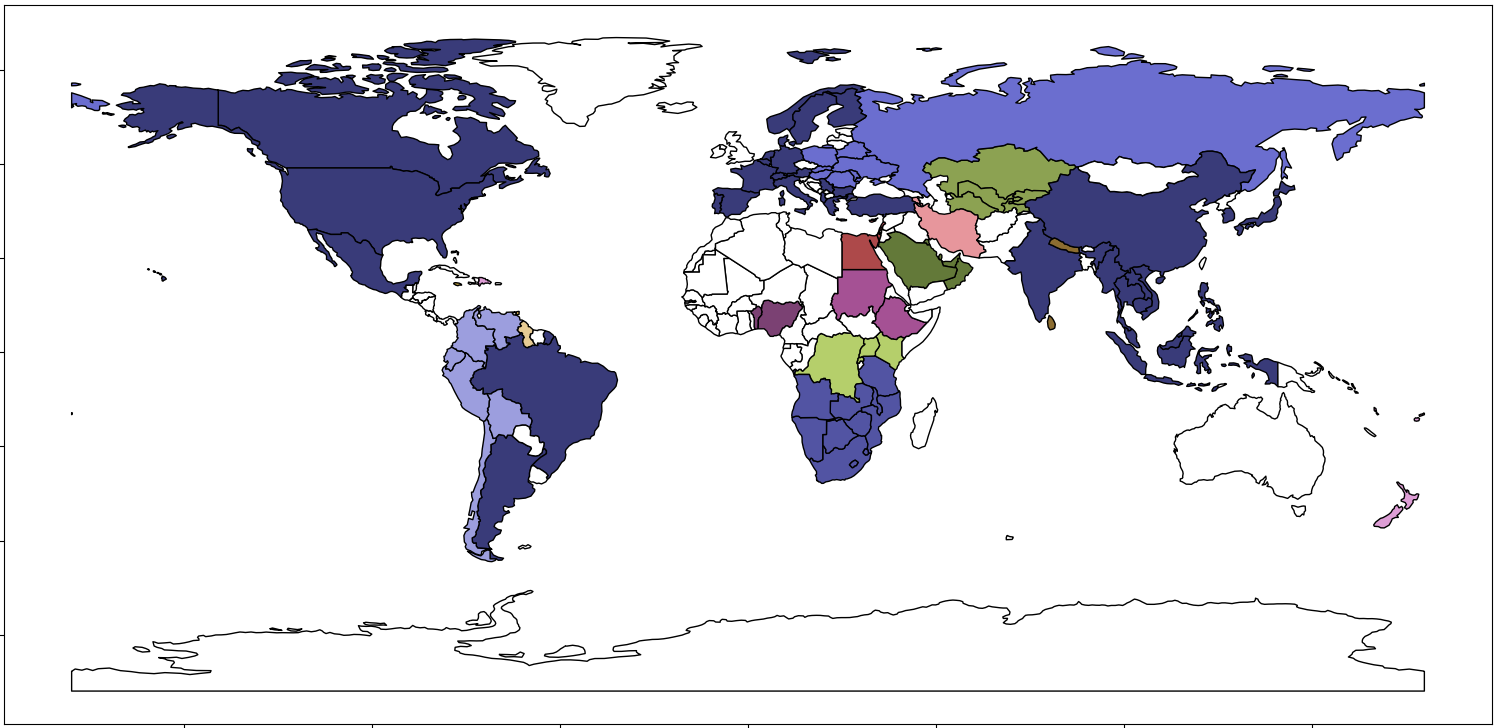}\label{fig_clusters_out_wto}}
 \caption{Clusters found for the Top-3 Out graph generated for the Foursquare and WTO datasets. Colors in each figure represent a specific cluster. Countries in white have not been included in the model or have been ignored because they are part of a single-country component/cluster.}
\label{figTop3OutComparison}
\end{figure*}

According to the procedure described in Section~\ref{sec:methodology}, clusters of countries can be obtained for both Top-k In and Top-k Out subgraphs. For comparison, Figures~\ref{fig_clusters_in} and~\ref{fig_clusters_in_wto} show the results for Top-3 In subgraphs of Foursquare and WTO, respectively. Figures~\ref{fig_clusters_out} and~\ref{fig_clusters_out_wto} show the results for Top-3 Out subgraphs of Foursquare and WTO, respectively.

Not surprisingly, most clusters are aligned with geographical regions. The complete list of countries in each cluster for Foursquare data is shown in Appendix \ref{appendix2}, and the list of countries in the WTO clusters can be found in \cite{lozano2018complex}.

As other studies have already observed \cite{hawelka2014geo,SILVA201795,belyi2017global,britoamcis18}, many clusters are formed in geographically close regions, even if physical distance is not one of the variables used to calculate the clustering. Furthermore, the clusters found with the Foursquare data are similar to those obtained based on the WTO data. There are some small divergences between the clusters found in this work and in \cite{lozano2018complex}. The observed exceptions are supposed to occur as we deal with two different data sources. This can be attributed to the lack of data for countries less central in international tourism and smaller Foursquare penetration.

Even with these divergences, the clusters are geopolitically consistent with what is expected, including geographically close countries and/or countries with affinities such as similar languages or cultural habits. Some examples from the top-3 Out subgraph are the cluster of south-west Asian/north African and majority Muslim countries (United Arab Emirates, Egypt, Kuwait, Qatar, Oman, Sudan, Bahrain, Pakistan, and Saudi Arabia) and the cluster of South-American countries (Peru, Argentina, Uruguay, Paraguay, Brazil, and Chile). Examples from the top-3 In subgraph are a cluster comprised of France, two of its overseas departments and two of its former colonies (France, Martinique, Guadeloupe, Tunisia, and Morocco), and a cluster of Baltic/southern Slavic countries (Bosnia and Herzegovina, Montenegro, and Serbia).

\subsection{Cross-regional Tourism Flow}
\label{sec:fluxCont}

To provide a macroscopic view of  Top-3 In and Top-3 Out subgraphs, Tables~\ref{continente_out} and~\ref{continente_in} describe the flow of tourism between continents instead of countries. 
The value in each cell is the number of different tourists traveling from any country in a specific continent (row) to any country in a particular continent (column). For example, the cell North America to South America in Table  \ref{continente_out} represents 332 trips of unique users from a country in North America to a country in South America.

\begin{table}[htbp]
\scriptsize
\caption{Cross-regional tourism flows for the Top-3 Out Foursquare subgraph. Cell values represent number of tourist's trips in a transition.}
\begin{center}
\begin{tabular*}{\tblwidth}{@{}LLLLLLL@{}}
\toprule
& \textbf{North} & \textbf{South} & \textbf{Europe} & \textbf{Africa} & \textbf{Asia} & \textbf{Oceania} \\
& \textbf{America} & \textbf{America} & & & \\
\midrule
\textbf{North America} & \textbf{6406} & 332 & 930 & 0 & 1 & 0 \\
\textbf{South America} & 1690 & \textbf{3104} & 0 & 0 & 0 & 0 \\
\textbf{Europe} & 566 & 0 & \textbf{9627} & 0 & 2461 & 0 \\
\textbf{Africa} & 46 & 0 & 99 & \textbf{23} & 204 & 0 \\
\textbf{Asia} & 560 & 0 & 3124 & 5 & \textbf{15989} & 0 \\
\textbf{Oceania} & 113 & 0 & 0 & 0 & 97 & \textbf{79} \\
\bottomrule
\end{tabular*}
\label{continente_out}
\end{center}
\end{table}

\begin{table}[htbp]
\scriptsize
\caption{Cross-regional tourism flows for the Top-3 In subgraph Foursquare. Cell values represent number of tourist's trips in a transition.}
\begin{center}
\begin{tabular*}{\tblwidth}{@{}LLLLLLL@{}}
\toprule
& \textbf{North} & \textbf{South} & \textbf{Europe} & \textbf{Africa} & \textbf{Asia} & \textbf{Oceania} \\
& \textbf{America} & \textbf{America} & & & \\
\midrule
\textbf{North Amer.} & \textbf{6862} & 1172 & 1917 & 50 & 1059 & 116 \\
\textbf{South Amer.} & 1215 & \textbf{3057} & 113 & 85 & 3 & 0 \\
\textbf{Europe} & 41 & 0 & \textbf{8266} & 269 & 1687 & 0 \\
\textbf{Africa} & 0 & 0 & 0 & \textbf{19} & 0 & 0 \\
\textbf{Asia} & 0 & 0 & 7093 & 458 & \textbf{17389} & 295 \\
\textbf{Oceania} & 0 & 0 & 0 & 0 & 0 & \textbf{48} \\
\bottomrule
\end{tabular*}
\label{continente_in}
\end{center}
\end{table}

As observed in \cite{lozano2018complex}, Tables ~\ref{continente_out} and~\ref{continente_in} make it clear that intracontinental tourism flows (i.e., tourism flows between countries of the same continent, represented by the values in the main diagonal) are greater than intercontinental ones. In addition, the small amount of data corresponding to tourists from Oceania and Africa is also confirmed in the WTO data, although in different proportions. 

\begin{figure}[htbp]
\centerline{\includegraphics[width=8cm]{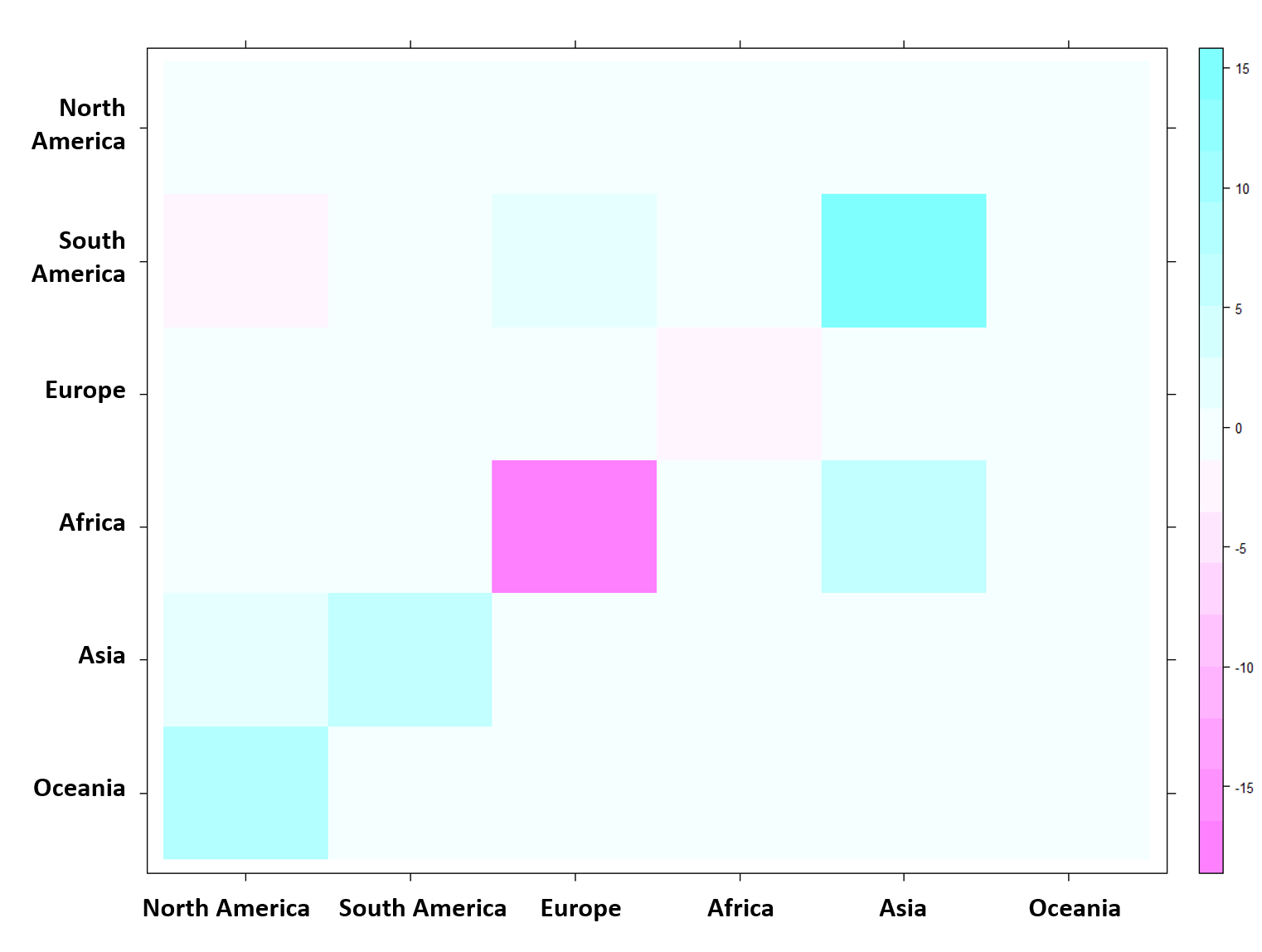}}
\caption{Percentage difference between Foursquare and WTO dataset (\cite{lozano2018complex}) of the intercontinental tourism flows for the Top-3 Out subgraphs. Negative values indicate lower values for Foursquare than WTO.}
\label{fig:continente_out_100}
\end{figure}

\begin{figure}[htbp]
\centerline{\includegraphics[width=8cm]{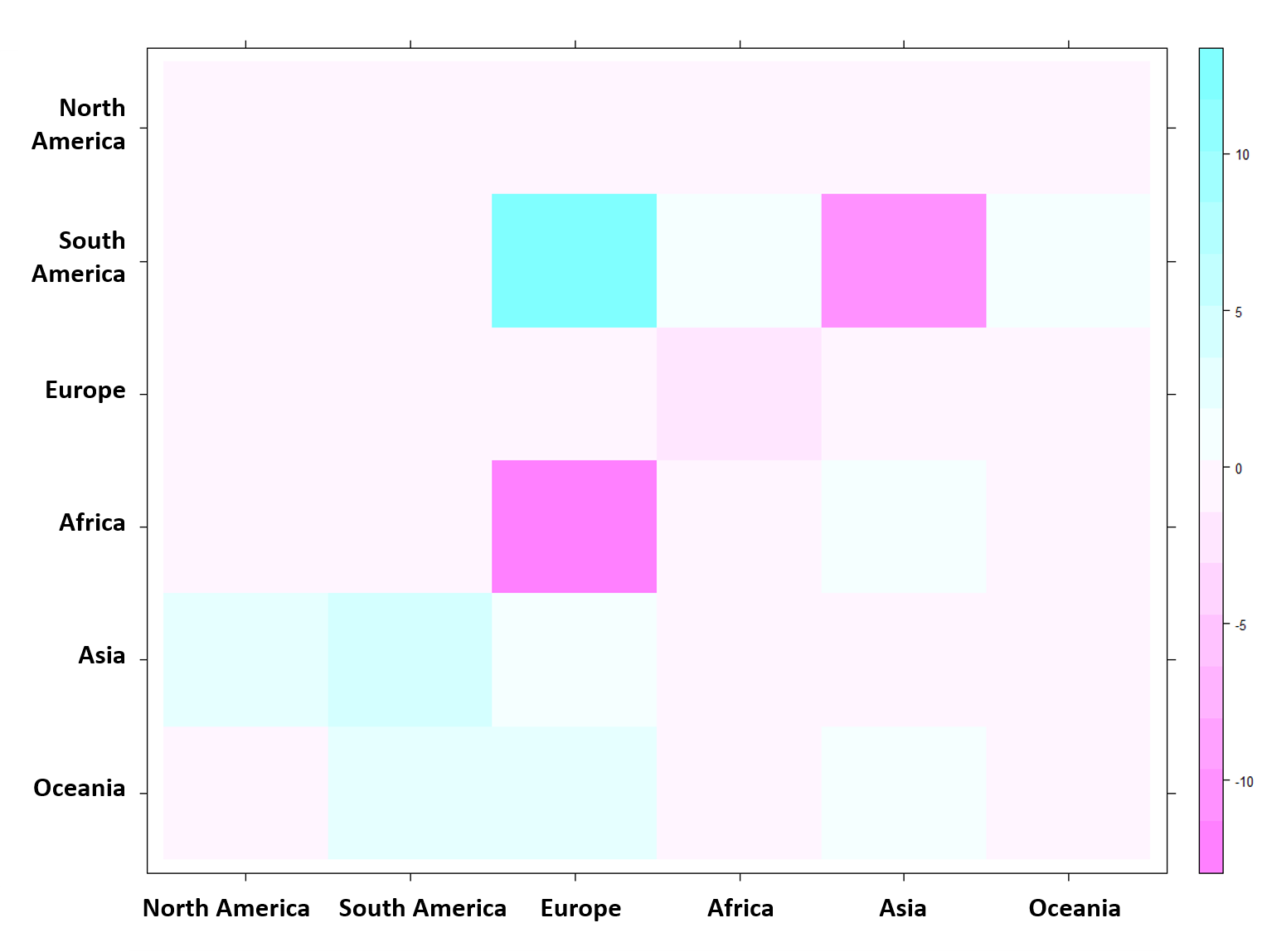}}
\caption{Percentage difference between Foursquare and WTO dataset (\cite{lozano2018complex}) of the intercontinental tourism flows for the Top-3 In subgraphs. Negative values indicate lower values for Foursquare than WTO.}
\label{fig:continente_in_100}
\end{figure}

A comparison between the data from both datasets is made by calculating the percentage difference between the flows of each possible transition. The results are shown in Figures~\ref{fig:continente_out_100} and~\ref{fig:continente_in_100}.

For Top-3 In subgraphs, the average percentage difference in the number of tourists is 1.52\%. For Top-3 Out, the average is  1.74\%. Due to many empty transitions (no tourist trips registered) for both datasets, there are several null cells. These small differences between the two datasets indicate that intercontinental relationships are well represented in the Foursquare dataset.

\subsection{Motif Analysis}
\label{sec:motifs}

Motif analysis identifies the most common edge patterns in the network, comparing their densities with the density of the same patterns in random nets of equal size. Table~\ref{tab:motifs} shows important motifs identified by mfinder using both datasets. The cells contain the Z-score of a pattern frequency relative to its frequency in a set of random graphs of the same size: To the left, data regarding the Foursquare dataset, and to the right, WTO data. Cells with NA represent patterns not identified as relevant for that subgraph and dataset.
 
\begin{table}[htbp]
\scriptsize
\caption{Three node motif analysis results. Each cell contains the Z-score of a pattern frequency relative to its frequency in a set of random graphs of the same size. The value to the left in a cell is associated with the Foursquare dataset, and the value to the right is associated with the WTO data.  Cells filled with NA represent patterns not identified as relevant for that subgraph and dataset.}
\begin{center}
\begin{tabular*}{\tblwidth}{@{}LLLLL@{}}
\toprule
Motif & Top-2 Out & Top-3 Out & Top-2 In & Top-3 In \\
\midrule
A$\leftarrow$B$\rightarrow$C, A$\rightarrow$C & 7.4/4.21 & 20.9/3.41  & 12.5/4.73  & 3.2/5.66  \\
A$\leftrightarrow$B$\rightarrow$C, A$\rightarrow$C & 0.8/NA  & 3.6/NA  & 4.2/7.06  & 12.1/10.38  \\
A$\leftarrow$B$\leftrightarrow$C, A$\rightarrow$C & NA/NA & 2.0/NA  & NA/5.31 & 12.1/NA  \\
A$\rightarrow$B$\leftrightarrow$C, A$\rightarrow$C & 2.0/7.19  & 9.5/4.15  & NA/NA & 3.9/5.8  \\
A$\leftrightarrow$B$\leftrightarrow$C, A$\rightarrow$C & NA/NA & 2.2/7.2  & 0.3/NA  & 2.0/15.09  \\
A$\leftrightarrow$B$\leftrightarrow$C, A$\leftrightarrow$C & NA/NA & 0.2/NA  & NA/NA & NA/NA \\
\bottomrule
\end{tabular*}
\label{tab:motifs}
\end{center}
\end{table}

There is a high correspondence between the most relevant \textit{motifs} of  Foursquare and WTO networks. The only pattern that has been identified as relevant for a Foursquare subgraph and not relevant for any WTO subgraphs is the complete triad (last row of Table~\ref{tab:motifs}). It may indicate a particular point that deserves further investigation in the future because it could be the case that LBSN data express certain relationships more precisely. Probably, it is complementary to other investigated sources, such as the WTO surveys investigated in the present work.

\begin{figure}[htbp]
\centerline{\includegraphics[width=8.5cm]{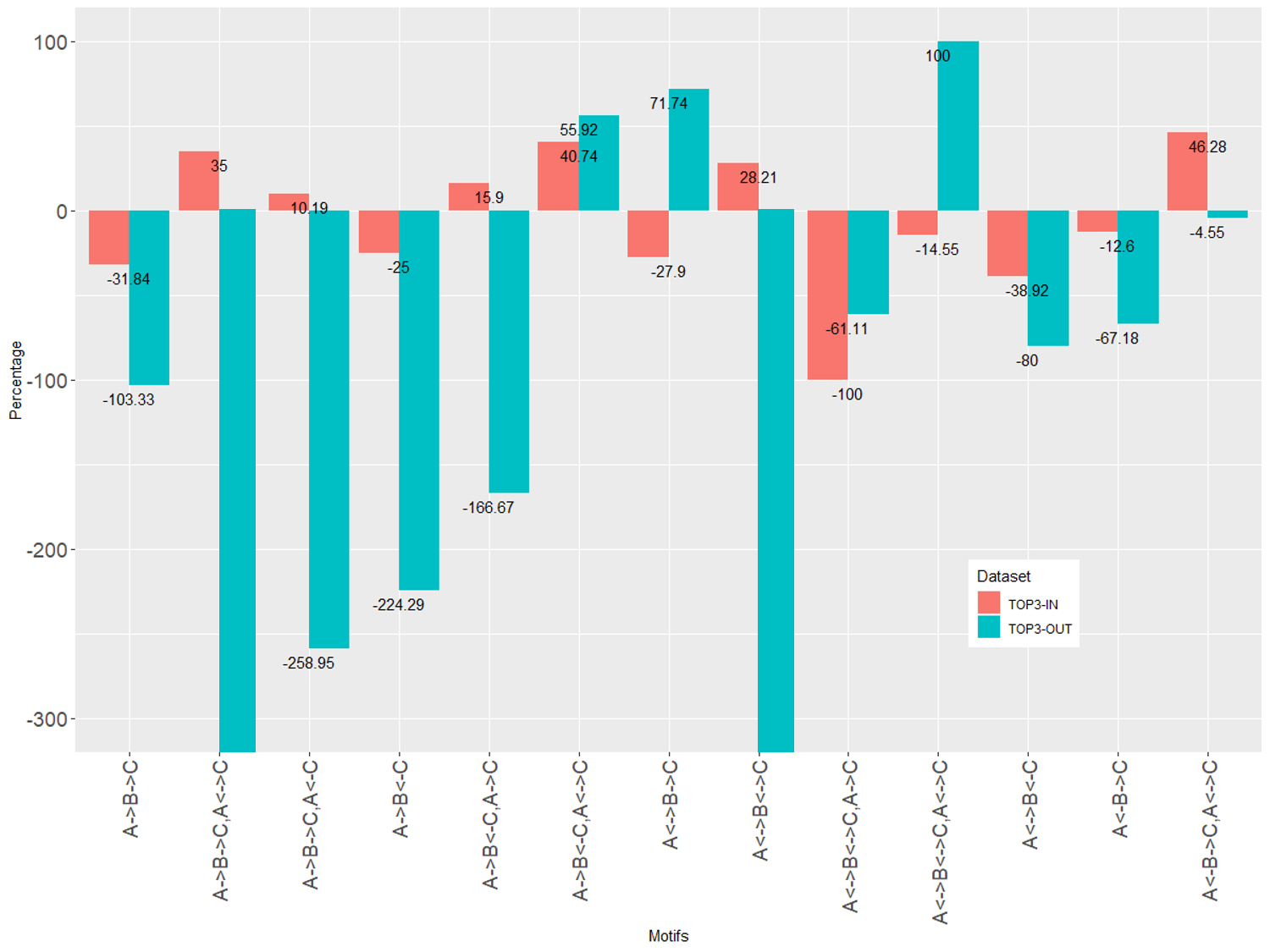}}
\caption{Percentage differences between the Z-values for each motif in each Top-3 subgraph for each dataset.}
\label{fig:comp_motifs}
\end{figure}

Figure~\ref{fig:comp_motifs} shows the percentage differences between the Z-values in the Top-3 In (red) and Out (blue) subgraphs of the two datasets for all three-node motifs (not only ones identified as relevant by mfinder in Table~\ref{tab:motifs}). Two difference values (-818.66 and -792.67) are not properly shown in Figure~\ref{fig:comp_motifs} due to scale limits. The average of the absolute values of the percentage differences is 32.86\% for the Top-3 In subgraph and 216.77\% for the Top-3 Out. These high differences may be due to bias in the LSBN data, favoring countries with a central role in international tourism whose residents/visitors have better access to the internet.

\subsection{Multidimensional Comparison}


So far,  we have observed, exploring individual metrics, considerable similarities between the two datasets under analysis, yet some dissimilarities have also been observed. In this section, we improve the previous analyses by considering several metrics together. For this multidimensional experiment, we aggregate all metrics that could be directly interpreted as features. Five metrics are common to Top-k In and Out subgraphs, k=\{1,2,3\}: In-strength, Out-strength, betweenness, pagerank, and ID of the strongly-connected component the country belongs to. Two are complementary for both types of subgraphs: In-degree for Top-k Out and Out-degree for Top-k In. Aiming to aggregate numerical and nominal metrics, in the preprocessing step we perform the one-hot encoding process on the nominal variable `component' and standardize all features by removing the mean and scaling to unit variance. 

Considering that we have  Top-(1,2,3) In and Out subgraphs, each one providing different metric values, every country is associated with six distinct multidimensional vectors - one for each subgraph. Therefore, for each country, an $n$-dimensional vector of features is built, associated to each subgraph. The dimension $n$ includes five elements for In-strength, Out-strength, betweenness, pagerank, and In-degree (or Out-degree) features plus as many elements as needed by the one-hot encoding. This last number of elements varies between 63 (in the case of Top-3 Out) to 107 (in the case of Top-1 In) in WTO data. By considering $m$ countries, a square $m$-dimensional symmetric matrix of distances is built with the euclidean distance $d_{ij}$ between the $n$-dimensional feature vectors  of countries $i$ and $j$. Therefore,  six matrices are obtained for all Top-$\{1,2,3\}$ In and Out subgraphs. An average matrix of distances is then computed element by element over these six matrices. Each row of the resulting matrix thus contains the average distance between one particular country and the others. Each dataset (Foursquare and WTO) yields to its own average matrix.

\begin{figure}[htbp]
\centerline{\includegraphics[width=8.7cm]{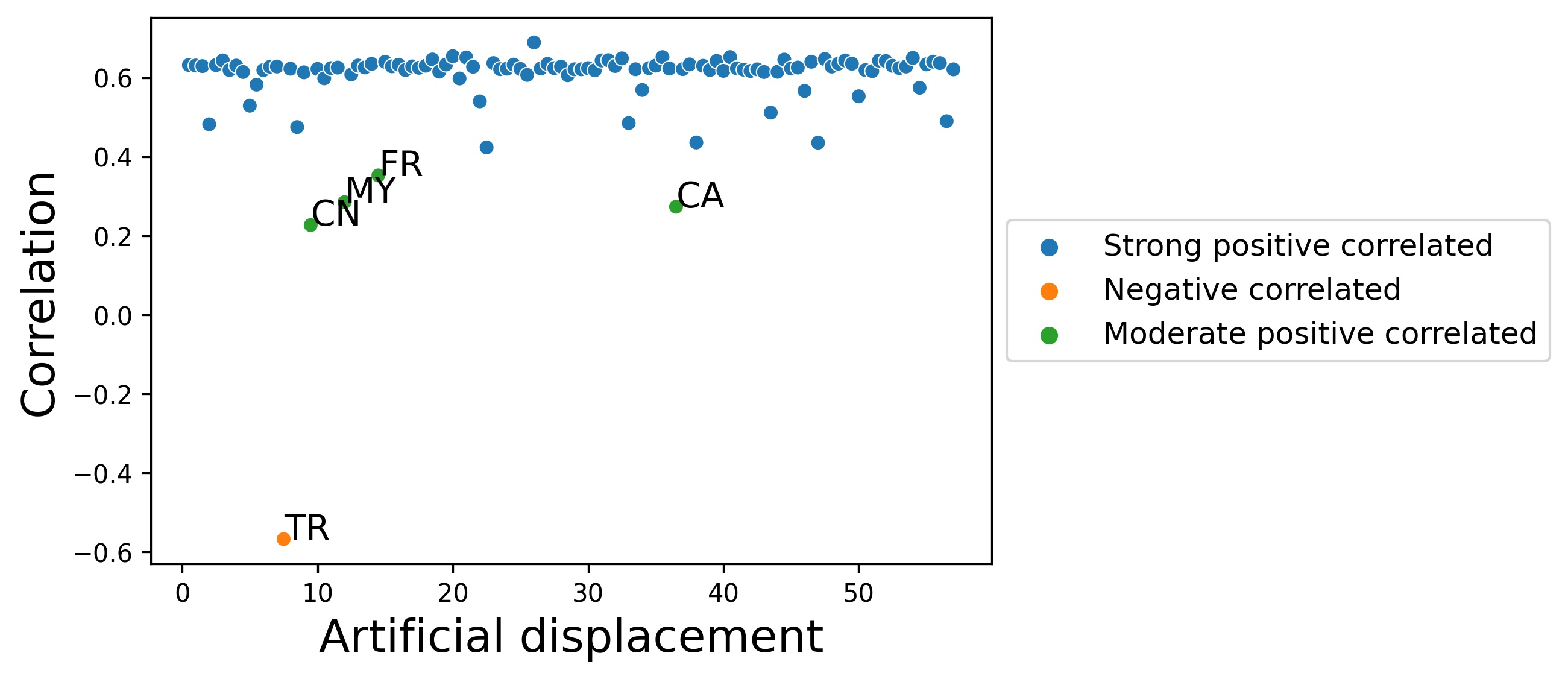}}
\caption{Pearson correlation coefficient expressing the relationship between distances from one country to all others using Foursquare and WTO. The x-axis, artificial displacement, is a controlled value just for visualization purposes; every point has one value that does not overlap another. }
\label{fig:correlations}
\end{figure}

The Pearson correlation $\rho$ for a particular country is then computed between its associated rows in Foursquare and in WTO average matrices. It represents how similar the country is related to the others in both datasets.
Figure \ref{fig:correlations} shows the results for all countries that appear in both datasets. First, note that for most of these countries, $\rho \geq 0.6$, confirming the indication of similarity we observed in the previous sections. The figure also highlights countries with $\rho \in [0.2,0.4]$, i.e., moderate positive correlations: Canada, China, France, Malaysia. One reason that might help explain these results is that those countries receive a high volume of tourists from different countries. Because social media data tend to be biased toward certain individuals in certain countries \cite{Silva:2019:UCL:3309872.3301284}, this bias might impact a comprehensive analysis. For instance,   France and Malaysia being more represented in the Foursquare dataset than their neighbors might also  affect their distances to other countries. In addition, the fact that China restricts the use of certain web services might also play an important role in this result. Canada's strong ties to the US across all subgraphs might skewer its results.  Only one result corresponding to Turkey has a negative correlation. As we observed in the results from the previous sections, Turkey's disproportionate popularity in the Foursquare dataset has influenced the results, impacting the analysis as well. Moreover, differently from what is observed in WTO data, few African countries appear in the Foursquare analysis. Therefore, unbalanced country information in each dataset is an important issue, even considering countries that appear in both datasets.

\section{Conclusions} \label{sec:conclusions}

According to the World Tourism Organization (WTO), the urban tourism industry has become essential to several economies, producing,  according to the WTO, more than US\$1 billion only in 2019. Another benefit is the ability of such an industry to generate and sustain millions of jobs and companies around the world. In this context, studying tourists and their behavior is essential to facilitate the growth and improvement of the urban tourism industry. The specific field of tourist mobility is still poorly studied, especially on large scale. The difficulty in building suitable data sets can be cited as one of the main reasons for this gap.

Traditional data sources for studying tourist movements, such as surveys, do not scale easily and sometimes do not provide necessary details. Therefore, there is a need to investigate alternative sources to study the phenomena. In this work, we identified that Location-Based Social Networks (LBSN), specifically Foursquare, are comparable to some extent to official data and can satisfactorily represent the reality of the global tourism network. This happens even with some known limitations of LBSNs, such as the predominance of some demographic groups - mainly young people with regular internet access \cite{Silva:2019:UCL:3309872.3301284}.

As expected, the results pointed out differences between survey and Foursquare data analyses. Many of the greatest differences occurred in the case of individual country analyses such as PageRank centrality. They differed mainly in the case of smaller and less central countries for international tourism, for which there are fewer data. Analyses considering the graph as a whole, such as the motif analysis, had highly similar results. A multidimensional comparison between the results obtained from Foursquare and WTO datasets showed that most countries have correlation values higher than $0.6$,  which means that both datasets share similar information. However, Canada, China, France, and Malaysia have moderated positive correlations between $0.2$ and $0.4$, and only Turkey has a negative correlation, probably due to the biased popularity of Turkey in Foursquare. These results suggest that Foursquare data can be useful for representing tourism behaviors with limitations due to the different popularity of social media in certain countries. This point should be considered in the future.

We would also like to point out that the strengths of LBSN data are not necessarily in macro analyses, thinking in terms of countries or regions as units. With these sources, it is possible to work at the level of individuals and/or specific addresses, going into more detail. 

The results obtained so far have opened up a range of new opportunities for expanding and complementing studies on tourist movement that rely on traditional sources, such as data from  WTO. With LBSN data, we can reach a global scale with a relatively low cost; in addition, we have the opportunity to explore a fine granularity in available data. Thus, for example, new applications and services could benefit from LBSN data to develop innovative solutions to meet the demands of the competitive global tourism.

\section*{Acknowledgements}

This study is also partially supported by the project GoodWeb (Grant 2018/23011-1 and 2020/07494-2 from S\~{a}o Paulo Research Foundation - FAPESP), and CNPq (grants  439226/2018-0, 314699/2020-1, 310998/2020-4).


\begin{thebibliography}{10}
\expandafter\ifx\csname url\endcsname\relax
  \def\url#1{\texttt{#1}}\fi
\expandafter\ifx\csname urlprefix\endcsname\relax\def\urlprefix{URL }\fi
\expandafter\ifx\csname href\endcsname\relax
  \def\href#1#2{#2} \def\path#1{#1}\fi

\bibitem{unwto20}
International Tourism Highlights, 2020 Edition, World Tourism Organization,
  Madrid, Spain, 2021.
\newblock \href
  {http://arxiv.org/abs/https://www.e-unwto.org/doi/pdf/10.18111/9789284422456}
  {\path{arXiv:https://www.e-unwto.org/doi/pdf/10.18111/9789284422456}}, \href
  {http://dx.doi.org/10.18111/9789284422456}
  {\path{doi:10.18111/9789284422456}}.

\bibitem{Zieba17}
M.~Zieba, Cultural participation of tourists--evidence from travel habits of
  austrian residents, Tourism Economics 23~(2) (2017) 295--315.

\bibitem{Scuderi18}
R.~Scuderi, C.~Dalle~Nogare, Mapping tourist consumption behaviour from
  destination card data: What do sequences of activities reveal?, International
  Journal of Tourism Research 20~(5) (2018) 554--565.

\bibitem{lozano2018complex}
S.~Lozano, E.~Guti{\'e}rrez, A complex network analysis of global tourism
  flowfs, International Journal of Tourism Research 20~(5) (2018) 588--604.

\bibitem{zheng2014}
Y.~Zheng, L.~Capra, O.~Wolfson, H.~Yang, Urban computing: Concepts,
  methodologies, and applications, ACM Trans. Intell. Syst. Technol. 5~(3)
  (2014) 38:1--38:55.

\bibitem{Silva:2019:UCL:3309872.3301284}
T.~H. Silva, A.~C. Viana, F.~Benevenuto, L.~Villas, J.~Salles, A.~Loureiro,
  D.~Quercia, Urban computing leveraging location-based social network data: A
  survey, ACM Comput. Surv. 52~(1) (2019) 17:1--17:39.

\bibitem{murphy2014}
J.~Murphy, M.~W. Link, J.~H. Childs, C.~L. Tesfaye, E.~Dean, M.~Stern,
  J.~Pasek, J.~Cohen, M.~Callegaro, P.~Harwood, Social media in public opinion
  research: report of the aapor task force on emerging technologies in public
  opinion research, American Association for Public Opinion Research, Deerfield
  (2014) 1--57.

\bibitem{baghal2021}
T.~Al~Baghal, A.~Wenz, L.~Sloan, C.~Jessop, Linking twitter and survey data:
  asymmetry in quantity and its impact, EPJ Data Science 10~(32) (2021) 1--20.

\bibitem{miguens2008travel}
J.~Migu{\'e}ns, J.~Mendes, Travel and tourism: Into a complex network, Physica
  A: Statistical Mechanics and its Applications 387~(12) (2008) 2963--2971.

\bibitem{zheng2009mining}
Y.~Zheng, L.~Zhang, X.~Xie, W.-Y. Ma, Mining interesting locations and travel
  sequences from gps trajectories, in: Proceedings of the 18th International
  Conference on World Wide Web, 2009, pp. 791--800.

\bibitem{lew2006modeling}
A.~Lew, B.~McKercher, Modeling tourist movements: A local destination analysis,
  Annals of tourism research 33~(2) (2006) 403--423.

\bibitem{fennell1996tourist}
D.~A. Fennell, A tourist space-time budget in the shetland islands, Annals of
  Tourism Research 23~(4) (1996) 811--829.

\bibitem{hawelka2014geo}
B.~Hawelka, I.~Sitko, E.~Beinat, S.~Sobolevsky, P.~Kazakopoulos, C.~Ratti,
  Geo-located twitter as proxy for global mobility patterns, Cartography and
  Geographic Information Science 41~(3) (2014) 260--271.

\bibitem{senefonteSocInfo2020}
H.~Senefonte, G.~Frizzo, M.~Delgado, R.~Luders, D.~Silver, T.~Silva, {Regional
  Influences on Tourists Mobility Through the Lens of Social Sensing}, in:
  Proc.\ of the International Conference on Social Informatics (SocInfo'20),
  Pisa, Italy, 2020.

\bibitem{FERREIRA2020240}
A.~P. Ferreira, T.~H. Silva, A.~A. Loureiro, Uncovering spatiotemporal and
  semantic aspects of tourists mobility using social sensing, Computer
  Communications 160 (2020) 240 -- 252.

\bibitem{belyi2017global}
A.~Belyi, I.~Bojic, S.~Sobolevsky, I.~Sitko, B.~Hawelka, L.~Rudikova,
  A.~Kurbatski, C.~Ratti, Global multi-layer network of human mobility,
  International Journal of Geographical Information Science 31~(7) (2017)
  1381--1402.

\bibitem{provenzano2018mobility}
D.~Provenzano, B.~Hawelka, R.~Baggio, The mobility network of european
  tourists: a longitudinal study and a comparison with geo-located twitter
  data, Tourism Review 73~(1) (2018) 28--43.

\bibitem{zhou2016structure}
M.~Zhou, G.~Wu, H.~Xu, Structure and formation of top networks in international
  trade, 2001--2010, Social Networks 44 (2016) 9--21.

\bibitem{leung2012social}
X.~Y. Leung, F.~Wang, B.~Wu, B.~Bai, K.~A. Stahura, Z.~Xie, A social network
  analysis of overseas tourist movement patterns in beijing: The impact of the
  olympic games, International Journal of Tourism Research 14~(5) (2012)
  469--484.

\bibitem{piazzi2011destinations}
R.~Piazzi, R.~Baggio, J.~Neidhardt, H.~Werthner, Destinations and the web: a
  network analysis view, Information Technology \& Tourism 13~(3) (2011)
  215--228.

\bibitem{hagberg2008exploring}
A.~Hagberg, P.~Swart, D.~S~Chult, Exploring network structure, dynamics, and
  function using networkx, Tech. rep., Los Alamos National Lab (LANL), Los
  Alamos, USA (2008).

\bibitem{reback2020pandas}
{Pandas Official Library}, \url{https://pandas.pydata.org/}, [Online; accessed
  20-Nov-2021] (2021).

\bibitem{brin1998anatomy}
S.~Brin, L.~Page, The anatomy of a large-scale hypertextual web search engine,
  Computer networks and ISDN systems 30~(1-7) (1998) 107--117.

\bibitem{newman2018networks}
M.~Newman, Networks, Oxford university press, 2018.

\bibitem{tan2016introduction}
P.-N. Tan, M.~Steinbach, V.~Kumar, Introduction to data mining, Pearson
  Education, 2016.

\bibitem{harris2020array}
C.~R. Harris, K.~J. Millman, S.~J. van~der Walt, R.~Gommers, P.~Virtanen,
  D.~Cournapeau, E.~Wieser, J.~Taylor, S.~Berg, N.~J. Smith, R.~Kern, M.~Picus,
  S.~Hoyer, M.~H. van Kerkwijk, M.~Brett, A.~Haldane, J.~F. del R{\'{i}}o,
  M.~Wiebe, P.~Peterson, P.~G{\'{e}}rard-Marchant, K.~Sheppard, T.~Reddy,
  W.~Weckesser, H.~Abbasi, C.~Gohlke, T.~E. Oliphant, Array programming with
  {NumPy}, Nature 585~(7825) (2020) 357--362.

\bibitem{scikit-learn}
F.~Pedregosa, G.~Varoquaux, A.~Gramfort, V.~Michel, B.~Thirion, O.~Grisel,
  M.~Blondel, P.~Prettenhofer, R.~Weiss, V.~Dubourg, J.~Vanderplas, A.~Passos,
  D.~Cournapeau, M.~Brucher, M.~Perrot, E.~Duchesnay, Scikit-learn: Machine
  learning in {P}ython, Journal of Machine Learning Research 12 (2011)
  2825--2830.

\bibitem{milo2002network}
R.~Milo, S.~Shen-Orr, S.~Itzkovitz, N.~Kashtan, D.~Chklovskii, U.~Alon, Network
  motifs: simple building blocks of complex networks, Science 298~(5594) (2002)
  824--827.

\bibitem{SILVA201795}
T.~H. Silva, P.~O.~V. de~Melo, J.~M. Almeida, M.~Musolesi, A.~A. Loureiro, A
  large-scale study of cultural differences using urban data about eating and
  drinking preferences, Information Systems 72~(Supplement C) (2017) 95 -- 116.

\bibitem{britoamcis18}
S.~A. de~Brito, A.~L. Baldykowski, S.~A. Miczevski, T.~H. Silva, Cheers to
  untappd! preferences for beer reflect cultural differences around the world,
  in: Proc.\ of Americas Conf.\ on Information Systems (AMCIS'18), New Orleans,
  USA, 2018.

\bibitem{tarjan1972depth}
R.~Tarjan, Depth-first search and linear graph algorithms, SIAM journal on
  computing 1~(2) (1972) 146--160.

\bibitem{nuutila1994finding}
E.~Nuutila, E.~Soisalon-Soininen, On finding the strongly connected components
  in a directed graph, Information processing letters 49~(1) (1994) 9--14.

\bibitem{borgatti2002ucinet}
I.-A. Apostolato, An overview of software applications for social network
  analysis., International Review of Social Research 3~(3) (2013) 1--9.

\end{thebibliography}



\appendix

\section{Countries Considered in this Study}\label{appendix1}

These are the countries/territories from each continent considered in the present study (in parenthesis, we show the number of countries/territories considered in relation to the total number in that continent):

\begin{itemize}
    \scriptsize
    \item \textbf{North America} (19 out of 45): the US, Mexico, Costa Rica, Canada, Dominican Republic, Martinique, Panama, El Salvador, Puerto Rico, Nicaragua, Guatemala, Honduras, Guadeloupe, Jamaica, Haiti, Aruba, Antigua and Barbuda, Trinidad and Tobago;
    \item \textbf{South America} (10 out of 15): Brazil, Chile, Paraguay, Colombia, Peru, Argentina, Venezuela, Ecuador, Uruguay, Bolivia;
    \item \textbf{Europe} (38 out of 49): Russia, UK, Spain, Latvia, Ukraine, Belgium, Germany, Italy, France, Netherlands, Cyprus, Belarus, Greece, Portugal, Hungary, Serbia, Poland, Sweden, Austria, Czech Republic, Finland, Romania, Switzerland, Ireland, Bulgaria, Denmark, Macedonia, Kosovo, Croatia, Vatican, Norway, Lithuania, Georgia, Estonia, Bosnia and Herzegovina, Slovakia, Slovenia, Montenegro;
    \item \textbf{Africa} (11 out of 55): Egypt, South Africa, Kenya, Morocco, Tunisia, Sudan, Uganda, Ghana, Nigeria, Cameroon, Côte d'Ivoire;
    \item \textbf{Asia} (37 out of 48): Turkey, Malaysia, Japan, Thailand, Indonesia, Philippines, Saudi Arabia, Kuwait, Singapore, South Korea, India, UAE, China, Sri Lanka, Hong Kong, Jordan, Lebanon, Taiwan, Maldives, Azerbaijan, Qatar, Oman, Kazakhstan, Vietnam, Pakistan, Bahrain, Israel, Iran, Kyrgyzstan, Uzbekistan, Brunei, Bangladesh, Palestine, Armenia, Iraq, Nepal,  Cambodia;
    \item \textbf{Oceania} (2 out of 14): Australia, New Zealand;
\end{itemize}

\section{Strongly Connected Components Found for Foursquare Data}\label{appendix3}

We use Tarjan’s algorithm \cite{tarjan1972depth} with Nuutila’s modifications \cite{nuutila1994finding} (default parameters of NetworkX library). For the Top-3 Out subgraph obtained with Foursquare data, the strongly connected components with more than one element are:

\begin{scriptsize}
     Argentine, Poland, Greece, Cyprus, Spain, Canada, Brazil, the UK, Italy, Paraguay, France, Chile, Mexico, Vatican, Germany, Russia, Ukraine, the US, Netherlands, Turkey, and Belgium; Denmark and Sweden; Belarus, Estonia, Lithuania, Latvia, and Finland; Panama and Costa Rica; Guatemala and El Salvador; Slovenia, Croatia, Montenegro, Serbia, and Bosnia and Herzegovina; South Korea, Thailand, Indonesia, Singapore, Malaysia, and Japan; Kazakhstan, Kyrgyzstan, and Uzbekistan; Jamaica, and Trinidad and Tobago; Oman, UAE, Saudi Arabia, Kuwait, and Bahrain; Palestine, Jordan, and Israel; China, Hong Kong, and the Philippines; Georgia, and Armenia; Guadeloupe and Martinique; the Maldives and Sri Lanka; Australia, and New Zealand; Uganda, and Kenya.
\end{scriptsize}

For Top-3 In for Foursquare data, they are:

\begin{scriptsize}
    Macedonia and Kosovo; Bosnia and Herzegovina, Serbia, and Montenegro; Sri Lanka and Maldives; Singapore, Indonesia, Thailand, Japan, Malaysia and the Philippines; Kyrgyzstan and Kazakhstan; El Salvador and Guatemala; Kuwait, UAE, and Saudi Arabia; Martinique, and Guadeloupe; Uganda and Kenya; Germany, Spain, Lithuania, Turkey, Ukraine, Latvia, France, Netherlands, Russia, Belarus, Belgium, Poland, United Kingdom, and Cyprus; Costa Rica and Panama; Colombia, Canada, Chile, Paraguay, Mexico, Argentine, Brazil, and the US.
\end{scriptsize}

\begin{figure}[htbp]
\centerline{\includegraphics[width=.49\textwidth]{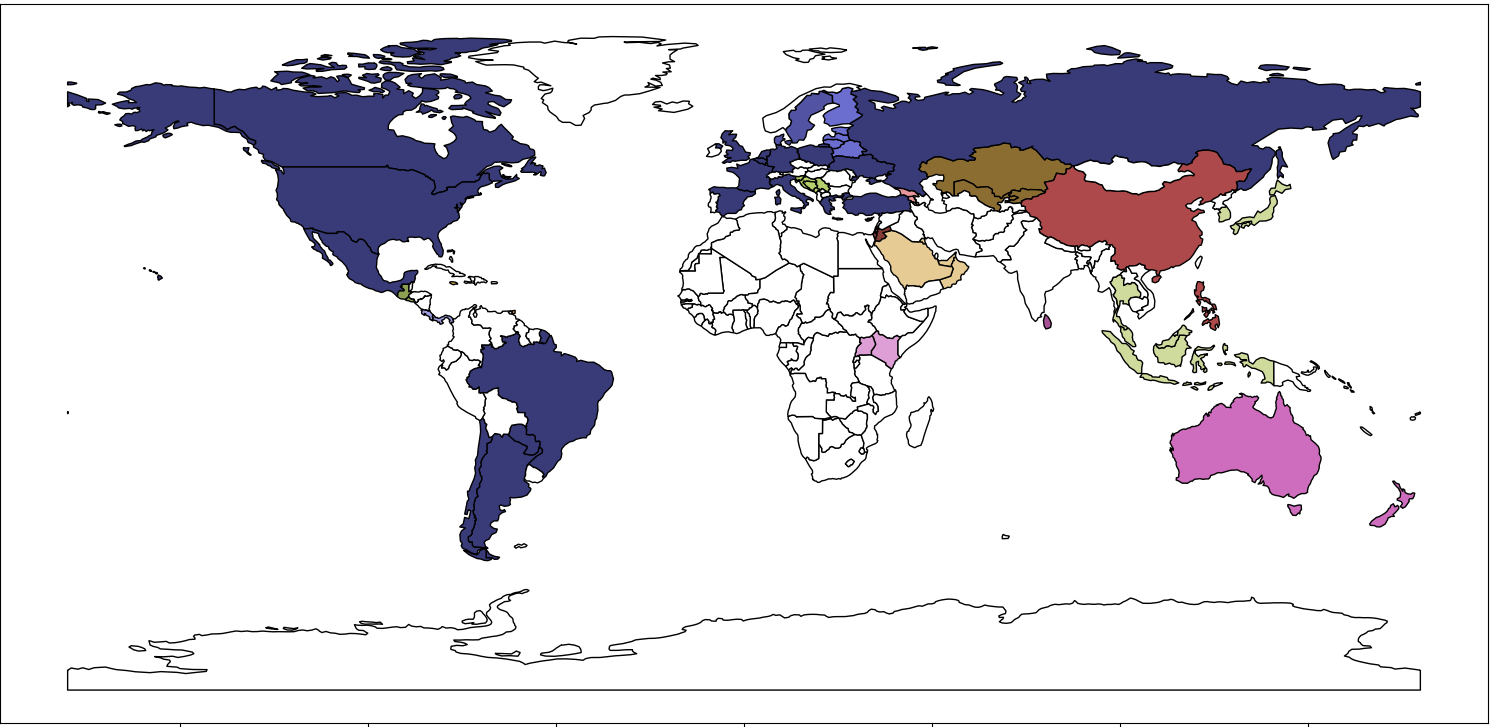}}
\caption{Strongly connected components in the Top-3 Out Foursquare subgraph. The same color indicates the same component. Countries in white have not been included in the model or have been ignored because they are part of a single-country component/cluster.}
\label{fig:components_out}
\end{figure}

\begin{figure}[htbp]
\centerline{\includegraphics[width=.49\textwidth]{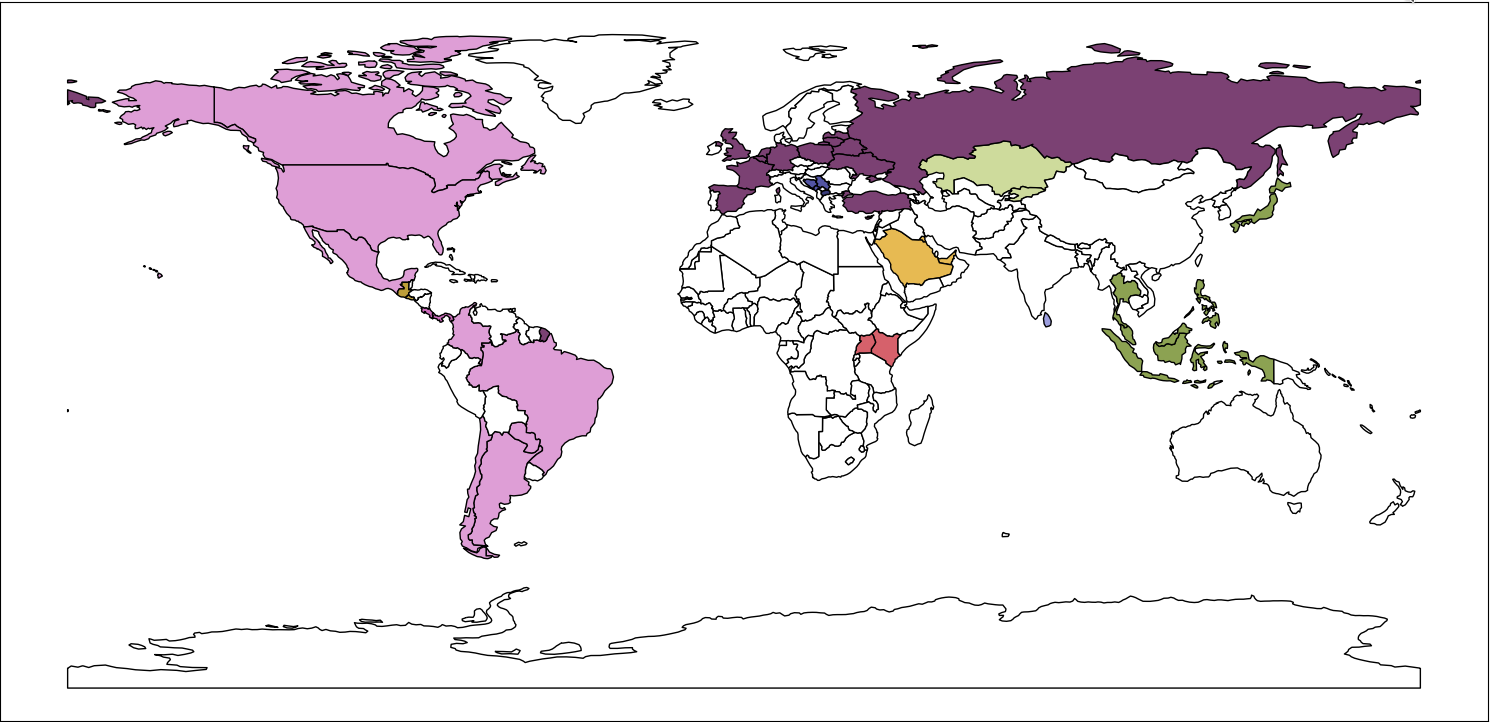}}
\caption{Strongly connected components in the Top-3 In Foursquare subgraph. The same color indicates the same component. Countries in white have not been included in the model or have been ignored because they are part of a single-country component/cluster.}
\label{fig:components_in}
\end{figure}

The components (ignoring those composed of only one country) for the Foursquare subgraphs are shown in Figures~\ref{fig:components_out} and~\ref{fig:components_in}.

\section{Clusters Found for Foursquare Data}\label{appendix2}

These are the clusters found with the hierarchical clustering algorithm for the Foursquare subgraphs:

Top-3 In subgraph clusters -- \begin{scriptsize} Mexico, the US, Puerto Rico, Dominican Republic, Costa Rica, Canada, Jamaica, Haiti, and Panama; the UK, Spain, Ireland, Norway, South Africa, and Portugal; Philippines and Hong Kong; Russia, Kazakhstan, Belarus, Ukraine, Uzbekistan, Latvia, and Lithuania;  France, Martinique, Guadeloupe, Tunisia, and Morocco; Trinidad and Tobago and Antigua, and Barbuda; Argentina, Uruguay, Paraguay, Brazil, Chile, and Bolivia; Bosnia and Herzegovina, Montenegro, and Serbia; Singapore, Malaysia, Brunei, Japan, Indonesia, Vietnam, Thailand, Taiwan, and South Korea;  Sweden, and Denmark;  Australia and New Zealand; Colombia, Venezuela, and Ecuador; Turkey, Georgia, Armenia, Kosovo, Romania, Iran, Cyprus, Azerbaijan, Northern Macedonia, Bulgaria, and Greece;  Uganda, and Kenya; Kuwait, Iraq, and Saudi Arabia;  El Salvador, and Guatemala; United Arab Emirates, Maldives, Oman, India, Pakistan, and Sri Lanka; Croatia, and Slovenia; Qatar and Sudan; Germany, and Austria;  Belgium, and the Netherlands.
\end{scriptsize}

Top-3 Out subgraph clusters -- \begin{scriptsize} Trinidad and Tobago, Mexico, Colombia, Venezuela, Ecuador, Aruba, Antigua and Barbuda, Bahamas, the US, Canada, and Jamaica; Singapore, Cambodia, Nepal, Malaysia, Japan, Bangladesh, Indonesia, Vietnam, Thailand, India and South Korea; United Arab Emirates, Egypt, Kuwait, Qatar, Oman, Sudan, Bahrain, Pakistan, and Saudi Arabia; Bosnia and Herzegovina, Kyrgyzstan, Turkey, Kazakhstan, and Cyprus; Finland, Georgia, Armenia, Norway, Russia, Sweden, Denmark, and Ukraine; Peru, Argentina, Uruguay, Paraguay, Brazil, and Chile; the UK, Côte d'Ivoire, France, Lebanon and Spain; Uganda, and Kenya; Martinique, and Guadeloupe; the Maldives, and Sri Lanka; Croatia, Slovenia, Vatican, Italy, Montenegro, and Serbia; Hungary, Slovakia, and Austria; Switzerland, Poland, Belgium, Czech Republic, Romania, Germany, and the Netherlands; Kosovo, Northern Macedonia, Bulgaria, and Greece; Palestine, Jordan, and Israel; Estonia, Belarus, Latvia, and Lithuania; Nigeria, and South Africa; Puerto Rico, and the Dominican Republic; China, the Philippines, Taiwan, and Hong Kong; Australia, and New Zealand; Nicaragua, Costa Rica, and Panama; Honduras, and El Salvador.
\end{scriptsize}

\section{Triad Census}\label{secTriadCensus}

\setcounter{table}{0}
\renewcommand{\thetable}{\Alph{section}.\arabic{table}}

Triad census consists of counting and classifying the patterns of edges between each set of three nodes in the graph. The analysis is done with UCINET 6.0 \cite{borgatti2002ucinet}, Table~\ref{censo_triades} shows the results of the census for Foursquare data. As mentioned in Section \ref{sec:methodology}, to compare the results of Foursquare with those provided by WTO, subgraphs density must be considered.

\begin{table}[ht]
\scriptsize
\caption{Triad census results. Number of occurrences of each pattern found in Top-k Out and Top-k In subgraphs.}
\begin{center}
\begin{tabular*}{\tblwidth}{@{}LLLLLLL@{}}
\toprule
\textbf{Triad} & \textbf{Top-1} & \textbf{Top-2} & \textbf{Top-3} & \textbf{Top-1} & \textbf{Top-2} & \textbf{Top-3}\\
& \textbf{Out} & \textbf{Out} & \textbf{Out} & \textbf{In} & \textbf{In} & \textbf{In}\\
\midrule
A,B,C & 248,425 & 237,931 & 228,246 & 248,187 & 237,330 & 227,622 \\
A$\rightarrow$B,C & 10,003 & 17,804 & 24,175 & 10,493 & 19,002 & 25,419 \\
A$\leftrightarrow$B,C & 1,217 & 3,083 & 5,255 & 743 & 1,826 & 3,469 \\
A$\leftarrow$B$\rightarrow$C & 0 & 23 & 82 & 604 & 1,556 & 2,720 \\
A$\rightarrow$B$\leftarrow$C & 390 & 933 & 1,592 & 0 & 33 & 71 \\
A$\rightarrow$B$\rightarrow$C & 47 & 85 & 150 & 41 & 111 & 189 \\
A$\leftrightarrow$B$\leftarrow$C & 48 & 188 & 411 & 0 & 6 & 22 \\
A$\leftrightarrow$B$\rightarrow$C & 0 & 11 & 39 & 62 & 199 & 399 \\ 
A$\rightarrow$B$\leftarrow$C,A$\rightarrow$C & 0 & 31 & 61 & 0 & 35 & 110 \\
A$\leftarrow$C,A$\rightarrow$C & 0 & 0 & 0 & 0 & 0 & 0 \\
A$\leftrightarrow$B$\leftrightarrow$C & 0 & 8 & 18 & 0 & 0 & 7 \\
A$\leftarrow$B$\rightarrow$C,A$\leftrightarrow$C & 0 & 17 & 50 & 0 & 4 & 20 \\
A$\rightarrow$B$\leftarrow$C,A$\leftrightarrow$C & 0 & 7 & 27 & 0 &21 & 59 \\
A$\rightarrow$B$\rightarrow$C,A$\leftrightarrow$C & 0 & 5 & 6 & 0 & 2 & 7 \\
A$\rightarrow$B$\leftrightarrow$C,A$\leftrightarrow$C & 0 & 4 & 13 & 0 & 5 & 13 \\
A$\leftrightarrow$B$\leftrightarrow$C,A$\leftrightarrow$C & 0 & 0 & 5 & 0 & 1 & 3 \\
\bottomrule
\end{tabular*}
\label{censo_triades}
\end{center}
\end{table}

The values distribution is similar between Foursquare and WTO data, especially regarding null elements.
A clear difference is the higher frequency of "complete" triads (A$\leftrightarrow$B$\leftrightarrow$C) on the Foursquare network. This could be only related to the difference in densities; however, the hypothesis that it is due to data's characteristics is reinforced since the pattern also emerges in the motif analysis.

\end{document}